\documentclass[review,3p]{elsarticle}

\usepackage{graphicx}
\usepackage{comment}
\usepackage{url}
\usepackage{multirow}
\usepackage{verbatim} 
\usepackage{booktabs}
\usepackage{amsmath}
\usepackage{amssymb}
\usepackage{algorithmic}
\usepackage[ruled]{algorithm2e}
\usepackage[utf8]{inputenc}

\usepackage{hyperref}
\hypersetup{
    colorlinks=true,
    linkcolor=blue,
    filecolor=magenta,      
    urlcolor=cyan,
    pdftitle={Overleaf Example},
    pdfpagemode=FullScreen,
    }

\renewcommand{\log}{\lg}
\newcommand{\bit}[1]{\texttt{#1}}

\newcommand{\rank}{\ensuremath{\mathsf{rank}}}
\newcommand{\select}{\ensuremath{\mathsf{select}}}
\newcommand{\access}{\ensuremath{\mathsf{access}}}
\newcommand{\bigoh}[1]{\ensuremath{O\!\left(#1\right)}}
\newcommand{\patrascu}{P\v{a}tra\c{s}cu\xspace}

\begin{document}


\title{Engineering Rank/Select Data Structures for Large-Alphabet Strings}



\author[dccUC,imfd]{Diego Arroyuelo\corref{corr}}
\ead{diego.arroyuelo@uc.cl}

\author[unipi]{Gabriel Carmona}
\ead{gabriel.carmona@phd.unipi.it}

\author[utfsm]{Héctor Larrañaga}
\ead{hector.larranaga@sansano.usm.cl}

\author[utfsm]{Francisco Riveros}
\ead{francisco.riverosc@sansano.usm.cl}

\author[utfsm]{Carlos Eugenio Rojas-Morales}
\ead{carlos.rojasmo@sansano.usm.cl}

\author[utfsm]{Erick Sep\'ulveda}
\ead{erick.sepulveda@alumnos.usm.cl}

\address[dccUC]{Department of Computer Science, Escuela de Ingeniería, Pontificia Universidad Católica de Chile.\\ Vicu\~na Mackenna 4860, Santiago, Chile}

\address[unipi]{Department of Computer Science, University of Pisa. Pisa, Italy}

\address[utfsm]{Department of Informatics, Universidad T\'ecnica Federico Santa Mar\'ia. Vicu\~na Mackenna 3939, Santiago, Chile}

\address[imfd]{Millennium Institute for Foundational Research on Data (IMFD)}


\cortext[corr]{Corresponding author.}


\begin{abstract}
Large-alphabet strings are common in scenarios such as information retrieval and natural-language processing. 
The efficient storage and processing of such strings usually introduces several challenges that are not witnessed in small-alphabets strings. This paper studies the efficient implementation of one of the most effective approaches for dealing with large-alphabet strings, namely the \emph{al\-pha\-bet-partitioning} approach. The main contribution is a compressed data structure that supports the fundamental operations $\rank$ and $\select$ efficiently.
We show experimental results that indicate that our implementation outperforms
the current realizations of the alphabet-partitioning approach.
In particular, the time for operation $\select$ can be improved by about 80\%,
using only 11\% more space than current alphabet-partitioning schemes.
We also show the impact of our data structure on several applications, like the intersection
of inverted lists (where improvements of up to 60\% are achieved, using only 2\% of extra space), the representation of run-length compressed strings, 
and the distributed-computation processing of $\rank$ and $\select$ operations.
In the particular case of run-length compressed strings, our experiments on the Burrows-Wheeler transform of highly-repetitive texts indicate that by using only about 0.98--1.09 times the space of state-of-the-art RLFM-indexes (depending on the text), the process of counting the number of occurrences of a pattern in a text can be carried out 1.23--2.33 times faster.
\end{abstract}



\maketitle


\section{Introduction}

Strings are a fundamental data type, intensively used in applications such as text, biological, and source code databases \cite{Navarro2016,Navjda13,MBCCNis16,TKKNIBTic22}. So, their efficient storage and manipulation is key. This means being able to (1) use as less space as possible to store them, and (2) manipulate them efficiently, supporting operations of interest. 
Let $s[1..n]$ be a string of length $n$, with symbols drawn from an alphabet 
$\Sigma = \{0,\ldots, \sigma-1\}$. We want to support the following operations on $s$, which will be the focus of this paper:
\begin{itemize}
\item $s.\rank_c(i)$: for $1\le i \le n$ and symbol $c\in\Sigma$, yields the number of occurrences of $c$ in $s[1..i]$;
\item $s.\select_c(j)$: for $c \in \Sigma$ and $1 \le j \le n_c$, yields the position of the $j$-th occurrence of symbol $c$ in $s$. Here, $n_c = s.\rank_c(n)$ is the total number of occurrences of $c$ in $s$; and
\item $s.\access(i)$: yields symbol $s[i]$, for $1 \le i \le n$.
\end{itemize}
In the particular case of binary strings $B[1{..}n]$ (or bit vectors), we have operations $B.\rank_{b}(i)$ and $B.\select_{b}(j)$, for $b \in \{\bit{0}, \bit{1}\}$.
Supporting operations $\rank$ and $\select$ is relevant for many applications \cite{Navarro2016},
such as snippet extraction in text databases \cite{AGMOSVis20},
query processing in information retrieval \cite{AGOspire10,AGOipm11}, and the space-efficient representation of
cardinal trees, text-search data structures, 
and graphs \cite{BCGNNalgo14}, among others.
As the amount of data processed by these applications is usually
large, we are interested in \emph{space-efficient} data structures to support these operations \cite{Navarro2016}. \emph{Succinct data structures} use space close to the
information-theory minimum, while supporting operations efficiently.
\emph{Compressed data structures}, on the other hand, take advantage
of certain regularities in the data to further reduce the space usage. The main motivation behind the use of space-efficient data structures is the in-memory processing of large amounts of data, avoiding expensive secondary-memory accesses, thus saving important time and improving energy consumption \cite{Fesa10}.


The main focus of this paper is the engineering of space-efficient data structures supporting $\rank$ and $\select$ operations on strings. The area of compact data structures is mature enough nowadays, with many theoretical challenges already solved \cite{AFScacm20}.  However, being able to properly implement the theoretical proposals is not trivial \cite{Navarro2016} and yields valuable results, usually requiring in-depth studies and strong algorithm engineering. 

We are particularly interested in large-alphabet strings, which are typical in scenarios like information retrieval and natural-language processing. Large alphabets introduce additional challenges when implementing $\rank/\select$ data structures. To illustrate this fact, notice the the most relevant data structures supporting $\rank$ and $\select$ operations on strings are \emph{wavelet trees} \cite{GGVsoda03}, using $n\lg{\sigma}(1+o(1)) + \Theta(\sigma w)$ bits of space, where $w = \Omega(\lg{n})$ is the word size (in bits) of the word-RAM model we assume in this paper (we shall use $\lg{x}$ to denote $\lceil\log_2{x}\rceil$). For large $\sigma$, the $\Theta(\sigma w)$-bit term (corresponding to wavelet tree pointers) can largely dominate the space usage. \emph{Wavelet matrices} \cite{CNPis15} tackle this linear alphabet dependency using just $n\lg{\sigma}(1+o(1))$ bits of space, while still supporting $\rank$, $\select$, and $\access$ in $\bigoh{\lg{\sigma}}$ time. The main idea is to concatenate all the bit vectors that represent the wavelet tree nodes into a single bit vector $\mathsf{wm}[1{..}n\lg{\sigma}]$ so as to replace wavelet tree nodes by operations $\rank$ and $\select$ on the bit vector. 
However, they have two main drawbacks: 
\begin{enumerate}
\item \textbf{The $\bigoh{\lg{\sigma}}$ computation time of $s.\rank$ and $s.\select$ can be high in practice}: although operations $\mathsf{wm}.\rank$ and $\mathsf{wm}.\select$ introduced to compute the wavelet tree pointers are supported in $O(1)$ time \cite{ClarkThesis,Mun96}, they are considerably more expensive than accessing a pointer. In particular, operation $\rank$ on a bit string is typically supported in $\sim$25--50 nanoseconds using the most efficient data structures from the \texttt{sdsl} library \cite{GPspe14}, whereas $\select$ typically takes $\sim$100--300 nanoseconds \cite{GPspe14,Kspire22}. The $\lceil\lg{\sigma}\rceil$ operations needed to navigate the tree can be high for large-alphabet string, hence the total operation time can become impractical. For instance, for the well-known GOV2 collection \cite{gov2}, $\lceil\lg{\sigma}\rceil = 25$, whereas for ClueWeb \cite{clueweb09} we have $\lceil\lg{\sigma}\rceil = 26$.

\item \textbf{The space usage is not compressed}: Wavelet matrices use space close to that of the plain representation of string $s$, no compression is achieved. In practice, strings have regularities that could be used to improve the space usage, using space proportional to some compressibility measure (such as, e.g., the empirical entropy of the string \cite{Man01}, or the number of equal-symbol runs in the string \cite{GNPjacm20}). Although a compressed bit vector can be used to represent $\mathsf{wm}$, like the one by Raman et al.~\cite{RRRtalg07}, this would increase the running time of $s.\rank$ and $s.\select$ considerably. 
\end{enumerate}

Another approach suitable for large alphabets is by Golynski et al.~\cite{GMRsoda06}, which improves the time of operation $\rank$ to $\bigoh{\lg\lg{\sigma}}$ and $\select$ to $\bigoh{1}$ time. This also improves the practical performance notoriously compared to wavelet matrices, yet still using non-compressed $n\lg{\sigma} + n\cdot o(\lg{\sigma})$ bits of space. 
The \emph{alphabet-partitioning} approach by Barbay et al.~\cite{BCGNNalgo14}, on the other hand, keeps essentially the same computation times as Golynsky et al.'s data structure, yet using compressed space, this way avoiding both drawbacks mentioned above for wavelet matrices. The main disadvantage is, however, that the theoretical times obtained by Barbay et al.~\cite{BCGNNalgo14} rely on the multi-ary wavelet trees by Ferragina et al.~\cite{FMMNtalg06} which are impractical according to a study by Bowe \cite{BoweThesis,Navarro2016}. 

\paragraph{Contributions}

In this paper, we study practical ways to implement the alphabet-partitioning approach with the main objective of improving the original
\cite{BCGNNalgo14} performance in practice. Our main contributions are as follows:
\begin{enumerate}
\item We carry out algorithm engineering on the \emph{alphabet-partition} approach by Barbay et al.~\cite{BCGNNalgo14}, obtaining an implementation that uses compressed space while supporting operations $s.\rank$ and $s.\select$ efficiently in practice. Our approach also yields interesting theoretical trade-offs, as it can be seen in Table \ref{tab:wt-state-of-the-art} (on page \pageref{tab:wt-state-of-the-art}).

\item We show that our approach yields competitive trade-offs when used for (i) snippet extraction from text databases and (ii) intersection of inverted lists. Both operations are of major interest for modern information retrieval systems \cite{BCC2010}.

\item We show that the alphabet-partition approach can be also used to improve run-length compression of large-alphabet strings formed by $r$ equal-symbol runs. In summary, we introduce a competitive alternative both in theory and practice. In particular, our idea allows us to introduce an integer parameter $c \ge 1$ such that the performance of the $\rank/\select$ data structure by Fuentes-Sepúlveda et al.~\cite{FKKPdcc18} can be expressed such as indicated in Table \ref{tab:wt-runs-state-of-the-art} (on page \pageref{tab:wt-runs-state-of-the-art}). We carry out extensive algorithm engineering to implement our run-length compression approach. In particular, we implement and try several data structures to represent bit strings with runs of \bit{0}s and \bit{1}s, which shall be used as building blocks for our data structures. We test our data structures for the classical problem of counting the number of occurrences of a pattern in a text, showing that state-of-the-art RLFM-indexes \cite{MNnjc05,GNPjacm20} can be improved by using our approach: by using only 0.98--1.09 times their space, counting the number of occurrences of a pattern in a text can be carried out 1.23--2.33 times faster (depending on the text).  

\item We show that our alphabet-partitioning scheme can be efficiently implemented on a distributed-memory system.
\end{enumerate}

An overall conclusion from our study is that our implementation of alphabet partitioning is not only effective (and efficient) to support the fundamental $\rank$ and $\select$ operations, but also to support several operations that are key for implementing modern information retrieval systems \cite{BCC2010,CB15}, one of the main applications that manipulate large-alphabet strings. The source code produced in this paper can be accessed at \url{https://github.com/Yhatoh/ASAP_paper/}.

\section{Related Work and Preliminary Concepts}
\label{sec:related}

The main problem we deal with in this paper is that of supporting operations $\rank$ and $\select$ efficiently on strings. We review in this section the main concepts and solutions for the problem.

\subsection{Succinct Data Structures for Bit Vectors}
\label{sec:succinct-data-structures}

Given a bit vector $B[1..n]$ with $m$ \texttt{1} bits, we want to support operations $B.\rank_{b}$ and $B.\select_{b}$, for $b \in \{\bit{0},\bit{1}\}$, as well as $B.\access$. The following space-efficient data structures offer the most interesting  trade-offs:
\begin{itemize}
\item The \texttt{SDarray} data structure by Okanohara and Sadakane \cite{OSalenex07} uses $m\log{\frac{n}{m}} + 2m + o(m)$ bits of space, and supports $\select$ in $O(1)$ time (provided we replace the $\rank$/$\select$ bit vector data structure on which \texttt{SDarray} is based by a constant-time $\select$ data structure \cite{Mun96}). Operations $\rank$ and $\access$ are supported in $O\left( \log{\frac{n}{m}}\right)$ time.

\item The data structure by Raman et al.~\cite{RRRtalg07} uses $\lg{n \choose m} + \bigoh{n/\lg^2{n}}$ bits of space, supporting $\rank$, $\select$, and $\access$ in $\bigoh{1}$ time.

\item The data structure by \patrascu \cite{Pfocs08} improves the second term in the space usage of Raman et al.'s data structure, requiring $\lg{n \choose m} + \bigoh{n/\lg^c{(n/c)}} + \bigoh{n^{3/4}\mathrm{polylog}(n)}$ bits of space, for any $c > 0$, while supporting all operations in $\bigoh{1}$ time. 

\item If the $m$ \bit{1}s of $B$ are grouped into $g$ runs, the data structure by Arroyuelo and Raman \cite{ARalgo22} uses at most $\lg{n-m+1 \choose g} + \lg{m-1 \choose g-1} + \bigoh{n/\lg^c{(n/c)}}$ bits of space, for any constant $c > 0$. Operation $\rank$ and $\select$ are supported in $\bigoh{1}$ time. It is important to note that $\lg{n-m+1 \choose g} + \lg{m-1 \choose g-1} \le \lg{n \choose m}$.
\end{itemize}

\subsection{Compressed Data Structures}

A \emph{compressed data structure} uses space proportional
to some compression measure of the data such as, e.g.,
the $0$-th order empirical entropy of a string $s[1..n]$ over an alphabet of
size $\sigma$, which is denoted by $H_0(s)$ and defined as:
\begin{equation}
\label{eq:h0}
H_0(s) =  \sum_{c\in\Sigma}{\frac{n_c}{n}\lg{\frac{n}{n_c}}},
\end{equation}
where $n_c$ is the number of occurrences of symbol $c$ in $s$. The sum includes
only those symbols $c$ that do occur in $s$, so that $n_c>0$.
The value $H_0(s)\le \lg{\sigma}$ is the average number of bits needed to encode
each string symbol, provided we encode them using $\lg{\frac{n}{n_c}}$ bits.

We can generalize this definition to that of $H_k(s)$, the $k$-th \emph{order empirical entropy} of string $s$, for $k > 0$, defined as
\begin{equation}
\label{eq:hk}
H_k(s) = \sum_{a \in \Sigma^k}{\frac{|s_a|}{n}H_0(s_a)},
\end{equation}
where $s_a$ denotes the string of symbols obtained from concatenating all symbols of $s$ that are preceded by a given context $a$ of length $k$. It holds that $0 \le H_k(s) \le \ldots \le H_1(s) \le H_0(s) \le \lg{\sigma}$, for any $k$.

Following Belazzougui and Navarro \cite{BNtalg15}, in general we call \emph{succinct} a string representation that uses $n\lg{\sigma} + o(n\lg{\sigma})$ bits, \emph{zeroth-order compressed} a data structure that uses $nH_0(s) + o(n\lg{\sigma})$ bits, and \emph{high-order compressed} one that uses $nH_k(s) + o(n\lg{\sigma})$ bits.  A \emph{fully-compressed} data structure is one such that the redundancy is also compressed, e.g., to use $nH_0(s)(1+o(1))$ bits.


\subsection{Rank/Select Data Structures on Strings}

Let $s[1{..}n]$ be a string of length $n$ with symbols drawn from an alphabet $\Sigma = \{0, \ldots, \sigma - 1\}$. We review next the state-of-the-art data structures for efficiently supporting $\rank$ and $\select$ on $s$. Table \ref{tab:wt-state-of-the-art} summarizes the space/time trade-offs of the approaches that we shall discuss next.

\begin{table}[ht]
    \begin{center}
    \caption{Space/time trade-offs for supporting operations $\rank$, $\select$, and $\access$ on a string $s[1{..}n]$ of symbols drawn from an alphabet of size $\sigma$.}
    \label{tab:wt-state-of-the-art}
\begin{tabular}{lcccc}
\toprule
Authors   & Space (bits) & \access & \rank & \select \\
\midrule
\cite{GGVsoda03}  &  $n\lg{\sigma} + o(n\lg{\sigma}) + \Theta(\sigma w)$ & \bigoh{\lg{\sigma}} & \bigoh{\lg{\sigma}} & \bigoh{\lg{\sigma}} \\
~\\
\cite{FMMNtalg06}, \cite{GRRswat08} &  $n(H_0(s) + o(1)) + \Theta(\sigma w)$                  & \bigoh{1+\frac{\lg{\sigma}}{\lg\lg{n}}} & \bigoh{1+\frac{\lg{\sigma}}{\lg\lg{n}}} & \bigoh{1+\frac{\lg{\sigma}}{\lg\lg{n}}} \\
~\\
\cite{CNPis15}  &  $n\lg{\sigma} + o(n\lg{\sigma})$ & \bigoh{\lg{\sigma}} & \bigoh{\lg{\sigma}} & \bigoh{\lg{\sigma}} \\
~\\
\cite{GMRsoda06}  & $n\lg{\sigma} + n\cdot o(\lg{\sigma})$ & \bigoh{\lg{\lg{\sigma}}} & \bigoh{\lg\lg{\sigma}} & \bigoh{1}\\
~\\
\cite{GORicalp10} & $nH_k(s) + \bigoh{\frac{n\lg{\sigma}}{\lg\lg{\sigma}}}$ (\dag) & \bigoh{1} & \bigoh{\lg\lg{\sigma}} & \bigoh{\lg{\lg{\sigma}}}\\
~\\
\cite{BCGNNalgo14} & $nH_0(s)(1+o(1))+o(n)$ & $\bigoh{\lg\lg{\sigma}}$ & $\bigoh{\lg\lg{\sigma}}$ & $\bigoh{1}$\\
~\\
\cite{BNtalg15}     & $n(H_0(s) + o(1))$ & $\bigoh{1+\frac{\lg{\sigma}}{\lg{w}}}$ & $\bigoh{1+\frac{\lg{\sigma}}{\lg{w}}}$ & $\bigoh{1+\frac{\lg{\sigma}}{\lg{w}}}$ \\ 
~\\
                    & $nH_0(s)(1+o(1))+o(n)$ & $\bigoh{1}$ & $\bigoh{\lg{\frac{\lg{\sigma}}{\lg{w}}}}$ & any $\omega(1)$ \\ 
~\\
                    & $nH_k(s) + \bigoh{n\lg\lg{\sigma}}$ (\ddag) & $\bigoh{1}$ & $\bigoh{\lg{\frac{\lg{\sigma}}{\lg{w}}}}$ & any $\omega(1)$ \\
\midrule
This paper & $nH_0(s)(1+o(1))+o(n)$ & $\bigoh{\max\{\lg^2{n}, L\}/L}$ $(\sharp)$ & $\bigoh{\lg\lg{\sigma}}$ & $\bigoh{1}$\\
~\\
           & $nH_0(s)(1+o(1))+o(n)$ & $\bigoh{\max\{\lg^2{n}, L\}/L}$ $(\sharp)$ & $\bigoh{\lg{\frac{\lg{\sigma}}{\lg{w}}}}$ & any $\omega(1)$\\

\bottomrule
\end{tabular}
\end{center}


(\dag): for any $k = o\left(\frac{\lg_\sigma{n}}{\lg\lg{\sigma}}\right)$.\\
(\ddag): for any $k = o\left(\lg_\sigma{n}\right)$, and $\lg{\sigma} = \omega(\lg{w})$.\\
$(\sharp)$: time per $\access$ for obtaining $L\ge 1$ consecutive symbols.
\end{table}

\subsubsection{Wavelet Trees}
A \emph{wavelet tree} \cite{GGVsoda03} (WT~for short) is a 
succinct data structure that supports operations $s.\rank$ and
$s.\select$ on a given string $s$, among many other operations (see, e.g., the work by Navarro \cite{Navjda13}).
The space usage is $n\lg{\sigma} + o(n\lg{\sigma}) + \Theta(\sigma w)$ bits \cite{GGVsoda03}, supporting $\rank$, $\select$, and $\access$ in $O(\lg{\sigma})$ time. The term $\Theta(\sigma w)$ in the space usage corresponds to the pointers needed to represent the $\Theta(\sigma)$-node binary tree on which a WT is implemented. This space is negligible for small alphabets, yet becomes dominant for large ones, e.g., $\sigma = \Theta(\sqrt{n})$ or extreme cases like $\sigma = \bigoh{n}$. There are practical scenarios where such alphabets are usual, e.g., in the representation of labeled graphs to support graph pattern matching \cite{AHNRRSsigmod21} and path queries \cite{AGHNRvldbj24}, or natural-language processing in information retrieval \cite{BCC2010}. 

To achieve compressed space, the WT can be given the shape of the Huffman tree, now using $nH_0(s)(1+o(1)) + \bigoh{n} + \Theta(\sigma w)$ bits. Operations take $O(\lg{n})$ worst-case, and $\bigoh{1+H_0(s)}$ time on average \cite{Navarro2016}.
Alternatively, one can use compressed bit vectors \cite{RRRtalg07} to represent each node of the original WT (i.e., not the Huffman-shaped WT). The space usage in that case is $nH_0(s) + o(n\lg{\sigma}) +\ \Theta(\sigma w)$ bits, whereas operations take $O(\lg{\sigma})$ time. Notice that even though one is able to compress the main components of the WT, the term $\Theta(\sigma w)$ keeps uncompressed and hence remains problematic for large alphabets.

The multi-ary WTs by Ferragina et al.~\cite{FMMNtalg06} support $\rank$ and $\select$ operations in $\bigoh{1+\frac{\lg{\sigma}}{\lg\lg{n}}}$ worst-case time, 
using $nH_0(S)+o(n\log{\sigma}) + \Theta(\sigma w)$ bits of space. Later, Golynski et al.~\cite{GRRswat08} improved the (lower-order term of the) space usage 
to $nH_0(s) + o(n) + \Theta(\sigma w)$ bits.
Notice that if $\sigma = \bigoh{\textrm{polylog}(n)}$, these approaches allow one to compute operations in $\bigoh{1}$ time.

To avoid the $\Theta(\sigma w)$ term in the space usage of WTs, \emph{wavelet matrices} \cite{CNPis15} concatenate the $\Theta(\sigma)$ WT nodes into a single bit vector $\mathsf{wm}[1{..}n\lg{\sigma}]$. The WT nodes are arranged within $\mathsf{wm}$ in such a way that the pointers between nodes can be simulated using operations $\mathsf{wm}.\rank$ (to go down to a child node) and $\mathsf{wm}.\select$ (to go up to a parent node) on bit vector $\mathsf{wm}$. Thus, operation $s.\rank$ and $s.\select$ on string $s$ are still supported in $\bigoh{\lg{\sigma}}$ time provided $\rank$ and $\select$ are supported in $\bigoh{1}$ time on the underlying bit vector \cite{ClarkThesis,Mun96}. However, even though WT pointers can be simulated in $\bigoh{1}$ time by using operations $\rank$/$\select$ on $\mathsf{wm}$, the constants associated to these operations are larger than the time of accessing a pointer. Hence, operations on wavelet matrices are slower in practice. The space usage is $n\lg{\sigma} + o(n\lg{\sigma})$ bits, avoiding the alphabet-unfriendly $\Theta (\sigma w)$-bit term. As wavelet matrices basically implement a WT avoiding the space of pointers, they are able to support also all operations a WT can support \cite{Navjda13}.

\subsubsection{Reducing to Permutations and Further Improvements}
Golynski et al.~\cite{GMRsoda06} introduce an approach that builds on data structures for permutations (and their inverses) \cite{MRRRicalp03}. The resulting scheme is more effective for larger alphabet than the above WTs schemes (including WM).
Their solution uses $n\lg{\sigma} + n\cdot o(\lg{\sigma})$ bits of space, supporting operation $\rank$ in $\bigoh{\lg\lg{\sigma}}$ time, and
operations $\select$ and $\access$ in $\bigoh{1}$ time (among other trade-offs, see the original paper for details). Later, Grossi et al.~\cite{GORicalp10} improved the space usage achieving high-order compression, that is, $nH_k(s) + \bigoh{\frac{n\lg{\sigma}}{\lg\lg{\sigma}}}$ bits. Operations $\rank$ and $\select$ are supported in $\bigoh{\lg\lg{\sigma}}$, whereas $\access$ is supported in $\bigoh{1}$ time. Finally, Belazzougui and Navarro \cite{BNtalg15} introduced optimal approaches, some of which are summarized in Table \ref{tab:wt-state-of-the-art}. There, ``any $\omega(1)$'' means any super-constant function $\omega(1)$ \cite{BNtalg15}.

\subsubsection{Alphabet Partitioning}
\label{sec:AP}

We are particularly interested in this paper in the \emph{alphabet-partitioning} approach \cite{BCGNNalgo14}, which is one of the most suitable for large alphabets.
Given a string $s[1{..}n]$ on alphabet $\Sigma = \{0, \ldots, \sigma-1\}$, the aim of alphabet partitioning
is to divide $\Sigma$ into $p$ smaller sub-alphabets $\Sigma_0, \Sigma_1, \ldots, \Sigma_{p-1}$,
such that $\bigcup_{i=0}^{p-1}{\Sigma_i} = \Sigma$, and 
$\Sigma_i\cap\Sigma_j = \emptyset$ for all $i\not = j$. Next, we discuss how this scheme is built to support $\rank$ and $\select$ on $s$.

\paragraph{The Mapping from Alphabet to Sub-Alphabet}
The data structure \cite{BCGNNalgo14} consists of an alphabet mapping
$m[1..\sigma]$ such that $m[i] = j$ iff
symbol $i\in \Sigma$ has been mapped to sub-alphabet $\Sigma_j$.
Within $\Sigma_j$, symbols are re-enumerated from 0 to $|\Sigma_j|-1$
as follows: if there are $k$ symbols smaller than $i$ in $\Sigma$
that have been mapped to $\Sigma_j$, then $i$ is encoded as $k$ in $\Sigma_j$.
Formally, $k = m.\rank_j(i)$.
From now on, let $n_j = |\{i\in [1{..}n], ~~m[s[i]] = j\}|$ denote the number of symbols of string $s$ that have been mapped to sub-alphabet $\Sigma_j$.
A way of defining the partitioning (which is called \texttt{sparse} \cite{BCGNNalgo14})
is:
\begin{equation}
m[\alpha] = \left\lceil \lg{\left(\frac{n}{n(\alpha)}\right)}\lg{n}\right\rceil,
\label{eq:sparse}
\end{equation}
where symbol $\alpha\in \Sigma$ occurs $n(\alpha)$ times in $s$.
Notice that $m[\alpha] \le \lceil \lg^2{n}\rceil$.

\paragraph{The Sub-Alphabet Strings}
For each sub-alphabet $\Sigma_\ell$, we store the subsequence $s_\ell[1..n_\ell]$, for $\ell =1, \ldots, p$, with the symbols of the original string $s$ that have been mapped to sub-alphabet $\Sigma_\ell$.


\paragraph{The Mapping from String Symbols to Sub-Alphabets}
\label{sec:defining-t}

In order to be able to reconstruct the original string, we store a sequence
$t[1..n]$, which maps every symbol $s[i]$ into the corresponding sub-alphabet $t[i]$.
That is, $t[i] = m[s[i]]$.
If $\ell = t[i]$, then the corresponding symbol
$s[i]$ has been mapped to sub-alphabet $\Sigma_\ell$,
and has been stored at position $t.\rank_\ell(i)$ in $s_\ell$.
Also, symbol $s[i]$ in $\Sigma$ corresponds to symbol
$m.\rank_\ell(s[i])$ in $\Sigma_\ell$.
Thus, we have $s_\ell[t.\rank_\ell(i)] = m.\rank_\ell(s[i])$.

\begin{figure}[ht]
$$s=  \mathtt{alabar\_a\_la\_alabarda}$$
$$\Sigma_0=\{\mathtt{a}\}; ~~~ \Sigma_1 = \{\mathtt{l}, \mathtt{\_}\}; ~~~ \Sigma_2 = \{\mathtt{b}, \mathtt{r}\}; ~~~ \Sigma_3 = \{\mathtt{d}\}$$

$$t= 1 ~~ 2 ~~ 1  ~~ 3 ~~ 1 ~~ 3 ~~ 2 ~~ 1 ~~ 2 ~~ 2 ~~ 1 ~~ 2 ~~ 1 ~~ 2 ~~ 1 ~~ 3 ~~ 1 ~~ 3 ~~ 4 ~~1$$

\begin{center}
\begin{tabular}{cccccccccc}
$s_0$ & & & $s_1$ & & & $s_2$ & & & $s_3$\\
$\mathtt{aaaaaaaaa}$ & & & $\mathtt{l\_\_l\_l}$ & & & $\mathtt{brbr}$ & & & $\mathtt{d}$  
\end{tabular}
\end{center}
\caption{Alphabet-partitioning data structure for the string $s=  \mathtt{alabar\_a\_la\_alabarda}$, assuming 4 sub-alphabets $\Sigma_0$, $\Sigma_1$, $\Sigma_2$, and $\Sigma_3$, the corresponding mapping $t$ and the sub-alphabet strings $s_0$, $s_1$, $s_2$, and $s_3$.} 
\label{fig:ap}
\end{figure}

Notice that $t$ has alphabet of size $p$. Also, there are $n_0$ occurrences of symbol 0
in $t$, $n_1$ occurrences of symbol 1, and so on. Hence, we define:
\begin{equation}
\label{eq:h0-t}
H_0(t) = \sum_{i=0}^{p-1}{\frac{n_i}{n}\lg{\frac{n}{n_i}}}.
\end{equation}

\paragraph{Computing the Operations}
Operations $\rank$, $\select$, and $\access$ are supported as follows, assuming that
$m$, $t$, and the sequences $s_\ell$ have been represented using appropriate $\rank$/$\select$
data structures (details about this later).
For $\alpha \in \Sigma$, let $\ell = m.\access(\alpha)$ be the partition containing $\alpha$ and let $c = m.\rank_\ell(\alpha)$ be the representation of symbol $\alpha$ in subalphabet $\Sigma_{\ell}$. Hence,
\begin{equation}\label{eq:rank-ap}
s.\rank_\alpha(i) \equiv s_\ell.\rank_c(t.\rank_\ell(i)),
\end{equation}
and
\begin{equation}\label{eq:select-ap}
s.\select_\alpha(j) \equiv t.\select_\ell(s_\ell.\select_c(j)).
\end{equation}
If we now define $\ell = t[i]$, then we have
\begin{equation}\label{eq:access-ap}
s.\access(i) \equiv m.\select_\ell(s_\ell.\access(t.\rank_\ell(i))).
\end{equation}

\paragraph{Space Usage and Operation Times}
Barbay et al.~\cite{BCGNNalgo14} showed that
$nH_0(t) + \sum_{\ell=0}^{p-1}{n_\ell \lg{\sigma_\ell}} \le nH_0(s) + o(n)$.
This means that if we use a zero-order compressed $\rank$/$\select$
data structure for $t$, and then represent every $s_\ell$ even in uncompressed form, we obtain zero-order compression for the input string $s$.
Recall that $p \le \lceil \lg^2{n} \rceil$, hence the alphabets of $t$ and $m$ are poly-logarithmic. Thus, a multi-ary wavelet tree \cite{GRRswat08} is used for $t$  and $m$, obtaining $O(1)$ time for $t.\rank$, $t.\select$, and $t.\access$ (equivalently for operations on $m$). The space usage
is $nH_0(t) + o(n)$ bits for $t$, and $O\left( \frac{n\lg\lg{n}}{\lg{n}}\right) H_0(s) = o(n)H_0(s)$ bits
for $m$.
For $s_\ell$, if we use Golynski et al.~data structure
\cite{GMRsoda06} we obtain a space usage of 
$n_\ell\lg{\sigma_\ell} + O\left( \frac{n_\ell \lg{\sigma_\ell}}{\lg\lg\lg{n}}\right)$ bits per partition,
and support operation $s_{\ell}.\select$ in $O(1)$ time, 
whereas $s_{\ell}.\rank$ and $s_{\ell}.\access$ are supported in $O(\lg\lg{\sigma})$ time.

Overall, the space is $nH_0(s) + o(n) (H_0(s) + 1)$ bits, operation
$s.\select$ is supported in $O(1)$ time, 
whereas operations $s.\rank$ and $s.\access$
are supported in $O(\lg\lg{\sigma})$ time
(see \cite{BCGNNalgo14} for details and further trade-offs).

\paragraph{Practical Considerations and Dense Partitioning}
In practice, the \texttt{sparse} partitioning defined in Equation (\ref{eq:sparse})
is replaced by an scheme such that for any $\alpha\in\Sigma$, $m[\alpha] = \lfloor \lg{r(\alpha)} \rfloor$ \cite{BCGNNalgo14}.
Here $r(\alpha)$ denotes the ranking of symbol $\alpha$ according to its frequency
(that is, the most-frequent symbol has ranking 1, and the least-frequent one has
ranking $\sigma$). Thus, the first partition contains only one symbol (the most-frequent
one), the second partition contains two symbols, the third contains four symbols, and so on.
Hence, there are $p=\lceil\lg{(\sigma+1)}\rceil$ partitions.
This approach is called \texttt{dense} \cite{BCGNNalgo14}.
Another practical consideration is to have a parameter $\ell_{min}$ for 
\texttt{dense}, such that the top-$2^{\ell_{min}}$ symbols in the ranking
are represented directly in $t$. That is, they are not represented in any
partition. Notice that the original \texttt{dense} partitioning can be achieved
by setting $\ell_{min} = 1$.

\subsection{Strings with Runs}

There are applications that must deal with strings $s$ formed by $r$ runs of equal symbols. Formally, let $s = c_1^{l_1}c_2^{l_2}\cdots c_r^{l_r}$, where:
\begin{itemize}
    \item $c_i \in \Sigma$, for $i=1,\ldots, r$, 
    \item $l_1, l_2, \ldots, l_r > 0$, 
    \item $c_i \not = c_{i+1}$ for all $1\le i < r$, and 
    \item $2\le \sigma \le r$. 
\end{itemize}
Interesting cases are strings with just a few long runs, as they are highly compressible. Typical examples are the Burrows-Wheeler transform \cite{BWT} of repetitive strings, such as those from applications like versioned document collections and source-code repositories.
Run-length encoding is the usual way to compress strings formed by runs, where $s$ is represented as the sequence $(c_1, l_1), (c_2, l_2), \ldots, (c_r, l_r)$. The space usage is $r(\lg{\sigma}+\lg{n})$ bits. Remarkable compression can be achieved in this way if $r$ is small (compared to the text length $n$). 

There are two main data structures in this case. First, the approach by M\"akinen and Navarro \cite{MNnjc05} uses $2r(2+\lg{(n/r)}) + \sigma\lg{n} + r\lg{\sigma}$ bits of space, supporting $\rank$ and $\access$ in $\bigoh{\lg{(n/r)} + \lg\lg{\sigma}}$, and $\select$ in $\bigoh{\lg{(n/r)}}$. On the other hand, Fuente-Sepúlveda et al.~\cite{FKKPdcc18} introduce a data structure using $(1+\epsilon)r\lg{\left(\frac{n\sigma}{r}\right)} + \bigoh{r}$ bits of space, for any $\epsilon > 0$. Operation $\rank$ is supported in $\bigoh{\lg{\frac{\lg{(n\sigma/r)}}{\lg\lg{n}}}}$, whereas $\select$ and $\access$ take $\bigoh{\lg{\frac{\lg{(n/r)}}{\lg\lg{n}}}}$ time. Table \ref{tab:wt-runs-state-of-the-art} summarizes these results. 

\begin{table}[ht]
    \begin{center}
    \caption{Comparison of rank/select data structures for strings with $r$ runs.}
    \label{tab:wt-runs-state-of-the-art}
\begin{tabular}{llccc}
\toprule
Authors   & Space (bits) & \access & \rank & \select \\
\midrule
\cite{MNnjc05} &  $2r(2+ \lg{\frac{n}{r}}) + \sigma\lg{n} + r\lg{\sigma}$ & $\bigoh{\lg{\frac{n}{r}} + \lg\lg{\sigma}}$ & $\bigoh{\lg{\frac{n}{r}} + \lg\lg{\sigma}}$ & $\bigoh{\lg{\frac{n}{r}}}$\\
~\\
\cite{FKKPdcc18} & $(1+\epsilon)r\lg{\frac{n\sigma}{r}}+\bigoh{r}$ & $\bigoh{\lg{\frac{\lg{(n/r)}}{\lg\lg{n}}}}$ & $\bigoh{\lg{\frac{\lg{(n\sigma/r)}}{\lg\lg{n}}}}$  & $\bigoh{\lg{\frac{\lg{(n/r)}}{\lg\lg{n}}}}$  \\
\midrule
This paper (\dag) & $(1+\epsilon)r^{(t)}\lg{\left(\frac{n\lg^c{n}}{r^{(t)}}\right)}+$  & $\bigoh{\lg{\frac{\lg{(n/r')}}{\lg\lg{n}}}}$ & $\bigoh{\lg{\frac{\lg{\frac{n\sigma}{r'\lg^c{n}}}}{\lg\lg{n}}}}$  & $\bigoh{\lg{\frac{\lg{(n/r')}}{\lg\lg{n}}}}$  \\
           & $+ (1+\epsilon)r^{(s)}\lg{\left(\frac{n\sigma}{r^{(s)}\lg^c{n}}\right)}+$ & \\
           & $+ \bigoh{r^{(t)} + r^{(s)}}$\\
\cmidrule{2-5}
& $2r^{(t)}\lg{\left(\frac{n}{r^{(t)}}\right)}+ r^{(t)}\cdot c\lg\lg{n}$  &   \\
           & $+ (1+\epsilon)r^{(s)}\lg{\left(\frac{n\sigma}{r^{(s)}\lg^c{n}}\right)}$ & 
           $\bigoh{A/ L}$  $(\sharp)$ & $\bigoh{\lg{\frac{\lg{\frac{n\sigma}{(r^{(s)}\lg^c{n})}}}{\lg\lg{n}}}}$ & $\bigoh{\lg{\frac{\lg{(n/r^{(s)})}}{\lg\lg{n}}}}$\\
           & $ + \bigoh{r^{(t)} + r^{(s)}}$\\
\bottomrule
\end{tabular}
\end{center}


(\dag): It holds $r^{(t)}\le r$, $r^{(s)} \le r$, see Section \ref{sec:run-strings} for their definitions, and $r' = \min{\{r^{(t)}, r^{(s)}\}}$. Here, $c\ge 1$ is any integer constant.\\
$(\sharp)$: time per $\access$ for obtaining $L\ge 1$ consecutive symbols, for $A = \max{
           \begin{cases}
           \lg{n}\\ L\cdot\lg{\frac{\lg{(n/r^{(s)})}}{\lg\lg{n}}}
           \end{cases}}$.
\end{table}

\section{A Faster Practical Alphabet-Partitioning Rank/Select Data Structure}

The alphabet-partitioning approach was originally devised to speed-up
decompression \cite{Said05}. Barbay et al.~\cite{BCGNNalgo14} showed that
alphabet partitioning is also effective for supporting operations 
$\rank$ and $\select$ on strings, being also one of the most
competitive approaches in practice.
In the original proposal \cite{BCGNNalgo14}, mapping $t$ (introduced in Section \ref{sec:defining-t})
is represented with a multi-ary WT \cite{FMMNtalg06,GRRswat08}, supporting $\rank$, $\select$,
and $\access$ in $O(1)$ time, since $t$ has alphabet of size $O(\textrm{polylog}(n))$.
In practice, Bowe \cite{BoweThesis} showed that multi-ary WTs can improve $\rank$/$\select$ computation time by about a half when compared to a binary WT \footnote{Operation times of up to 1 microsecond were shown by Bowe \cite{BoweThesis}.}, yet increasing the space usage noticeably ---making them impractical, in particular for representing sequence $t$. 
Consequently, in the experiments of Barbay et al.~\cite{BCGNNalgo14} no multi-ary WT was tested. Besides, the \texttt{sdsl} library \cite{GPspe14} uses a Huffman-shaped WT by default for $t$ in the template of class $\texttt{wt\_ap<>}$ \footnote{Actually, no multi-ary wavelet tree is implemented in the \texttt{sdsl}.}. Thus, operations $\rank$ and $\select$ on $t$ are not as efficient as one would expect in practice, requiring, e.g., hundreds of nanoseconds. 
We propose next an implementation of the alphabet-partitioning approach that avoids mapping $t$, storing the same information in a data structure that can be queried faster in practice while being space-efficient: just a few tens of nanoseconds, still using space close to $nH_0(t)$ bits. 
Our simple implementation decision has several important consequences in practice, not only improving the computation time of $\rank$/$\select$, but also that of applications such as inverted list intersection, full-text search on highly-repetitive text collections using compressed suffix arrays \cite{FMjacm05,GNPjacm20}, and the distributed computation of $\rank$/$\select$ operations. 


\subsection{Data Structure Definition}

Our scheme consists of the mapping $m$ and the sub-alphabets subsequences $s_0, \ldots, s_{p-1}$, one per partition, just as originally defined in Section \ref{sec:AP}. This time, however, we disregard mapping $t$, replacing it by bit vectors
$B_0[1{..}n], B_1[1{..}n], \ldots, B_{p-1}[1{..}n]$,  one per partition. See Figure \ref{fig:example-asap} for an illustration. For any partition $\ell = 0, \ldots, p-1$, we set $B_\ell[i] = \mathtt{1}$, for all $1\le i\le n$, iff $s[i] \in \Sigma_\ell$
(or, equivalently, it holds that $m.\access(s[i]) = \ell$). 
Notice that $B_{\ell}$ has $n_{\ell}$ \texttt{1}s. We represent these bit vectors using a data structure supporting operations $B_{\ell}.\rank_\mathtt{1}$ and $B_{\ell}.\select_\mathtt{1}$.
By using these bit vectors, we aim at improving the time spent on $t$ in practice for the original scheme, hence decreasing the time to compute operations
$s.\rank$ and $s.\select$. We will also show that by properly representing these bit vectors, the original space bounds can be also retained. 
%
%
\begin{figure}[ht]
    \centering
$$s=  \mathtt{alabar\_a\_la\_alabarda}$$
$$\Sigma_0=\{\mathtt{a}\}; ~~~ \Sigma_1 = \{\mathtt{l}, \mathtt{\_}\}; ~~~ \Sigma_2 = \{\mathtt{b}, \mathtt{r}\}; ~~~ \Sigma_3 = \{\mathtt{d}\}$$

\begin{center}
$B_0:  \mathtt{1  0  1  0  1  0  0  1  0  0  1  0  1  0  1  0  1  0  0  1}$\\
$B_1:  \mathtt{0  1  0  0  0  0  1  0  1  1  0  1  0  1  0  0  0  0  0  0}$\\
$B_2:  \mathtt{0  0  0  1  0  1  0  0  0  0  0  0  0  0  0  1  0  1  0  0}$\\
$B_3:  \mathtt{0  0  0  0  0  0  0  0  0  0  0  0  0  0  0  0  0  0  1  0}$\\
\end{center}

\begin{center}
\begin{tabular}{cccccccccc}
$s_0$ & & & $s_1$ & & & $s_2$ & & & $s_3$\\
$\mathtt{aaaaaaaaa}$ & & & $\mathtt{l\_\_l\_l}$ & & & $\mathtt{brbr}$ & & & $\mathtt{d}$  
\end{tabular}
\end{center}
    \caption{Our implementation of the alphabet-partitioning data structure for the string $s=  \mathtt{alabar\_a\_la\_alabarda}$, assuming 4 sub-alphabets $\Sigma_0$, $\Sigma_1$, $\Sigma_2$, and $\Sigma_3$. The original mapping $t$ is replaced by bit vectors $B_0$, $B_1$, $B_2$, and $B_3$.}
    \label{fig:example-asap}
\end{figure}

Operations $s.\rank_\alpha$ and $s.\select_\alpha$ are computed on our scheme as follows. 
Let $\alpha \in \Sigma$ be a symbol that has been mapped to sub-alphabet 
$\ell = m.\access(\alpha)$. Let $c=m.\rank_\ell(\alpha)$ be its representation in $\Sigma_\ell$. Hence, we define:
\begin{equation}\label{eq:rank-asap}
s.\rank_\alpha(i) \equiv s_\ell.\rank_c(B_\ell.\rank_\mathtt{1}(i)),
\end{equation}
and:
\begin{equation}\label{eq:select-asap}
s.\select_\alpha(j) \equiv B_\ell.\select_\mathtt{1}(s_\ell.\select_c(j)).
\end{equation}

A drawback is that operation $s.\access(i)$ cannot be supported
efficiently on our scheme: 
we do not know the partition $j$ such that $B_j[i] = \mathtt{1}$, as we do not know symbol $s[i]$ (indeed, that is the symbol we are asking for). 
Hence, for $j = 0, \ldots, p-1$, we must check $B_j[i]$, until for a given $\ell$ it holds that
$B_\ell[i] = \mathtt{1}$. This takes $O(p) = O(\log^2{n})$ time.
Then, we compute:
\begin{equation}\label{eq:access-asap}
s.\access(i) \equiv m.\select_\ell(s_\ell.\access(B_\ell.\rank_\mathtt{1}(i))).
\end{equation}
Although $\access$ cannot be supported efficiently, 
there are still relevant applications where this operation is not needed,
such as computing the intersection of inverted lists \cite{AGOspire10,BCGNNalgo14,Navarro2016},
or computing the term positions for phrase searching and positional ranking functions \cite{AGMOSVis20}.
Besides, many applications need operation $\access$ to obtain not just a single symbol, but a string snippet $s[i..i+L-1]$ of length $L$ ---e.g., snippet-generation tasks \cite{AGMOSVis20}. This operation can be implemented efficiently on our scheme as in Algorithm \ref{alg:snippet}. The idea is to obtain the snippet $s[i..i+L-1]$ by extracting all the relevant symbols from each partition. To do so, for every partition $\ell = 0,\ldots, p-1$, the number of $\mathtt{1}$s within $B_{\ell}[i{..}i+L-1]$ (which can be computed as $B_\ell.\rank_\mathtt{1} (i+L-1)- B_\ell.\rank_\mathtt{1} (i-1)$) indicates the number of symbols of $s_\ell$ that belong to the snippet $s[i..i+L-1]$. Also, the position of each such $\mathtt{1}$ within $B_\ell[i..i+L-1]$ corresponds to the position of the corresponding symbol of $s_\ell$ within the snippet. The main idea is to obtain these $\mathtt{1}$s using $\select_{\mathtt{1}}$ on $B_{\ell}$. 
\begin{algorithm}[ht]
\centering
\caption{$\mathsf{snippet}(i, L)$}
\label{alg:snippet}
\begin{algorithmic}[1]
\STATE Let $S[1..L]$ be an array of symbols in $\Sigma$.
\FOR{$\ell=0$ \TO $p-1$} 
\STATE $cur \gets B_\ell.\rank_\mathtt{1} (i-1)$
\STATE $count \gets B_\ell.\rank_\mathtt{1} (i+L-1)-cur$
\FOR{$k=1$ \TO $count$} 
\STATE $cur \gets cur +1$
\STATE $S[B_\ell.\select(cur)-i+1] \gets m.\select_j(s_j.\access(cur))$
\ENDFOR
\ENDFOR
\RETURN $S$
\end{algorithmic}
\end{algorithm}


\subsection{Space Usage and Operation Time}

To represent bit vectors $B_0, \ldots, B_{p-1}$, recall that we need to support $\rank_{\mathtt{1}}$ and $\select_{\mathtt{1}}$. The first alternative would be a plain bit vector representation, such as the one by Clark and Munro \cite{Mun96,CMsoda96,ClarkThesis}. Although operations $\rank$ and $\select$ are supported in $\bigoh{1}$ time, the total space would be $p\cdot n + o(p\cdot n) = \bigoh{n\lg^2{n}}$ bits, which is prohibitive. A more space-efficient alternative would be the data structure by \patrascu \cite{Pfocs08}, able to represent a given $B_\ell[1{..}n]$ (that has $n_\ell$ $\mathtt{1}$s) using $\lg{n \choose n_{\ell}} + \bigoh{n/\lg^{c}{(n/c)}} + \bigoh{n^{3/4}\mathrm{poly}\lg{n}}$ bits of space, for any $c > 0$. The total space for the $p$ bit vectors would hence be
$$\sum_{\ell=0}^{p-1}{\left(\lg{n \choose n_{\ell}} + \bigoh{\frac{n}{\lg^{c}{(\frac{n}{c})}}} + \bigoh{n^{3/4}\mathrm{poly}\lg{n}}\right)} =
\left(\sum_{\ell=0}^{p-1}{\lg{n \choose n_{\ell}}}\right) + p\cdot\left(\bigoh{\frac{n}{\lg^{c}{(\frac{n}{c})}}} + \bigoh{n^{3/4}\mathrm{poly}\lg{n}}\right).$$
For the first term, we have that
$$\sum_{\ell=0}^{p-1}{\lg{n \choose n_{\ell}}} \le nH_0(t),$$
whereas the second term can be made $o(n)$ by properly choosing a constant $c>2$ (since $p = \bigoh{\lg^2{n}}$). The total space for $B_0, \ldots, B_{p-1}$ is, thus, $nH_0(t) + o(n)$, supporting $\rank_\mathtt{1}$ and $\select_{\mathtt{1}}$ in $\bigoh{1}$ time \cite{Pfocs08}. We are able, in this way, to achieve the same space and time complexities as the original schema \cite{BCGNNalgo14}, replacing the multi-ary wavelet tree that represents $t$ by bit vectors $B_{0},\ldots, B_{p-1}$. Operations $s.\rank$ and $s.\select$ can be supported within the same bounds as for the original alphabet partitioning approach (see Table \ref{tab:wt-state-of-the-art}). Operation $\mathsf{snippet}$ (see Algorithm \ref{alg:snippet}) can be supported in $\bigoh{\lg^2{n} + L} = \max{\{\lg^2{n}, L\}}$ time. This means that if the snippet to be extracted is long enough (i.e., $L \ge \lg^2{n}$), the extraction of every symbol amortizes to $\bigoh{1}$ time. Alternatively, subsequences $s_0, \ldots, s_p$ can be represented with Belazzougui and Navarro \cite{BNtalg15} approach (in particular, their second trade-off we show in Table \ref{tab:wt-state-of-the-art}). In this case, we are able to improve the time for $s.\rank$ operation to $\bigoh{\lg{\frac{\lg{\sigma}}{\lg{w}}}}$, while slightly worsening $s.\select$'s performance to $\omega (1)$ time. See Table \ref{tab:wt-state-of-the-art} for the details of this trade-off.

A drawback about using \patrascu's data structure to represent $B_0, \ldots, B_{p-1}$ is, however, its impracticality: (1) there is no known practical implementation of this approach; (2) in practice, the term $p\cdot (\bigoh{n/\lg^c{(n/c)}} + \bigoh{n^{3/4}\mathrm{poly}\lg{n}})$ in the space usage is likely to be dominant and excessive. 
A more practical approach would be Raman et al.~\cite{RRRtalg07} approach, whose implementation is included in several libraries, such as \texttt{sdsl} \cite{GPspe14}. However, the additional $o(\cdot)$ term of this alternative would not be $o(n)$ as before and would need a non-negligible amount of additional space in practice. 

A third approach, more practical than  previous ones, is to represent the bit vectors $B_{\ell}$ using the \texttt{SDarray} representation by Okanohara and Sadakane \cite{OSalenex07},
requiring
$$\sum_{i=0}^{p-1}{n_i\lg{\frac{n}{n_i}} + 2n_i + o(n_i)}$$
bits of space, overall. Notice that for the first term we have
$\sum_{i=0}^{p-1}{n_i\log{\frac{n}{n_i}}} = nH_0(t)$, 
according to Equations (\ref{eq:h0}) and (\ref{eq:h0-t}).
Also, we have that $2\sum_{i=0}^{p-1}{n_i} = 2n$.
Finally, for $\sum_{i=0}^{p-1}{o(n_i)}$ we have that
each term in the sum is actually $O(n_i/\lg{n_i})$ \cite{OSalenex07}.
In the worst case, we have that 
every partition has $n_i = n/p$ symbols.
Hence, $n_i/\lg{n_i} = n/(p\lg{\frac{n}{p}})$, which for
$p$ partitions yields a total space of $O( n/\lg{\frac{n}{p}})$ bits. This is
$o(n)$ since $\lg{\frac{n}{p}} \in \omega(1)$ for $n \in \omega(p)$ (which is a reasonable assumption). In our case, $p \le \lg^2{n}$,
hence $\sum_{i=0}^{p-1}{o(n_i)} \in o(n)$.
Summarizing, bit vectors $B_\ell$ require
$n(H_0(t) + 2 + o(1))$ bits of space.
This is $2$ extra bits per symbol when compared to mapping $t$ from Barbay et al.'s approach \cite{BCGNNalgo14}.
The whole data structure uses $nH_0(s) + 2n + o(n)(H_0(s) + 1)$ bits.
Regarding the time needed for the operations, $s.\select$ 
can be supported in $O(1)$ time.
Operation $s.\rank$ can be supported in $O(\log{n})$ worst-case time:
if $n_i = O(\sqrt{n})$, operation
$B_i.\rank$ takes $O(\log{\frac{n}{n_i}}) = O(\log{n})$ time.

Using \texttt{SDArray}, Algorithm \ref{alg:snippet} ($\mathsf{snippet}$) takes 
$O\left(\sum_{i=0}^{p-1}{\lg{\frac{n}{n_i}}} + L\lg{\lg{\sigma}} \right)$ time.
The sum $\sum_{i=0}^{p-1}{\lg{\frac{n}{n_i}}}$ is maximized when
$n_i = n/p = n/\lg^2{n}$.
Hence, $\sum_{i=0}^{p-1}{\lg{\frac{n}{n_i}}} = \lg^2{n}\cdot \lg\lg{n}$,
thus the total time for $\mathsf{snippet}$ 
is $O\left( \lg^2{n}\cdot \lg\lg{n} + L\lg{\lg{\sigma}} \right)$. 
Thus, for sufficiently long snippets, this algorithm is faster than using
$\access$.
Regarding construction time, bit vectors $B_\ell$ can be constructed in linear time:
we traverse string $s$ from left to right; for each symbol $s[j]$, determine its partition
$\ell$ and push-back the corresponding symbol in $s_\ell$, as well as position $j$ into
an auxiliary array $A_\ell$, $0\le \ell \le p-1$. Afterwards, positions stored in $A_\ell$ corresponds to the positions of the \bit{1}s within $B_\ell$, so they can be used to construct the corresponding \texttt{SDarray}.

\subsection{Experimental Results on Basic Operations} \label{sec:experimental-setup}

In this section, we experimentally evaluate the performance of our approach to support the basic operations $\rank$, $\select$, and $\access$.
We implemented our data structure following the \texttt{sdsl} library \cite{GPspe14}.
Our source code was compiled using \texttt{g++} with flags \texttt{-std=c++11} and optimization flag \texttt{-O3}.
Our source code can be downloaded from our code repository \cite{our-code}.
We run our experiments on an HP Proliant server running an Intel(R) Xeon(R) CPU E5-2630 at 2.30GHz, 
with 6 cores, 15 MB of cache, and 384 GB of RAM.

As input string in our test we use a 3.0 GB prefix of the Wikipedia (dump from August 2016). 
We removed the \texttt{XML} tags, leaving just the text. 
The resulting text length is 505,268,435 words and the vocabulary has 8,468,328 distinct words. 
We represent every word in the text using a 32-bit unsigned integer, resulting in 1.9 GB of space.
The zero-order empirical entropy of this string is 12.45 bits.

We tested \texttt{sparse} and \texttt{dense} partitionings, the latter
with parameter $\ell_{min}=1$ (i.e., the original \texttt{dense} partitioning),
and $\ell_{min} = \lg{\lg{\sigma}} = \lg{23}$
(which corresponds to the partitioning scheme currently implemented in \texttt{sdsl}).
The number of partitions generated is 476 for \texttt{sparse}, 24 for \texttt{dense} $\ell_{min}=1$,
and 46 for \texttt{dense} $\ell_{min} = \lg{23}$.
For operations $\rank$ and $\select$,
we tested two alternatives for choosing the symbols $\alpha$ used for our $s.\rank_{\alpha}$ and $s.\select_{\alpha}$ queries:
\begin{itemize}
\item \textbf{Random symbols}: 30,000 vocabulary words generated by choosing positions of the input string uniformly at random, then using the corresponding word for the queries. This is the same approach used in the experimental study by Barbay et al.~\cite{BCGNNalgo14}.

\item \textbf{Query-log symbols}: we use words from the query log TREC 2007 Million Query Track 
\footnote{\url{http://trec.nist.gov/data/million.query/07/07-million-query-topics.1-10000.gz}}.
We removed stopwords, and used only words that exist in the vocabulary. 
Overall we obtained 29,711 query words (not necessarily unique).
\end{itemize}

For operation $s.\rank_{\alpha}(i)$, we generate position $i$ uniformly at random in the interval $[1{..}n]$.
For $s.\select_{\alpha}(j)$, we generate $j$ uniformly at random in the interval $[1{..}n_\alpha]$, where $n_{\alpha}$ is the number of occurrences of symbol $\alpha$ in $s$.
For operation $s.\access(i)$, we generate position $i$ uniformly at random in the interval $[1{..}n]$.

We call \texttt{ASAP} our approach, keeping the name of the original publication \cite{ASspire19}.
We test the following combinations for mapping $m$ and sequences $s_\ell$, as well as
\texttt{sparse} (\texttt{S}) and \texttt{dense} (\texttt{D}) partitioning \footnote{We only show the most competitive combinations we tried.}:
\begin{itemize}
\item \texttt{ASAP GMR-WM (D 23)}: the scheme using Golynski et al.~data structure \cite{GMRsoda06} (\texttt{gmr\_wt<>} in
\texttt{sdsl}) for $s_\ell$, using samplings 4, 8, 16, 32, and 64 for the inverse permutation data structure. For mapping $m$, this scheme uses a wavelet matrix (\texttt{wm\_int<>} in \texttt{sdsl}), and \texttt{dense} $\ell_{min} = \lg{23}$ partitioning.

\item \texttt{ASAP GMR-WM (D)}: the same approach as before,
this time using the original \texttt{dense} partitioning. 

\item \texttt{ASAP WM-AP (S)}: we use wavelet matrices for $s_{\ell}$ and alphabet partitioning for $m$, with \texttt{sparse} partitioning. 

\item \texttt{ASAP WM-HUFF\_INT (D 23)}: we use wavelet matrices for $s_{\ell}$ and a Huffman-shaped wavelet tree for $m$ (\texttt{wt\_huff<>} in \texttt{sdsl}). The partitioning is \texttt{dense} $\ell_{min} = \lg{23}$.

\item \texttt{ASAP WM-WM (S)}: we use wavelet matrices for both $s_{\ell}$ and $m$, with \texttt{sparse} partitioning.
\end{itemize}
Bit vectors $B_\ell$ are implemented using template $\texttt{sd\_vector<>}$ from \texttt{sdsl}, which corresponds to Okanohara and Sadakane's \texttt{SDArray} data structure \cite{OSalenex07}.

We compare with the following competitive approaches in the \texttt{sdsl}:
\begin{itemize}
\item \texttt{AP}: the original alphabet
partitioning data structure \cite{BCGNNalgo14}.
We used the default scheme from \texttt{sdsl},
which implements mappings $t$ and $m$ using Huffman-shaped WTs, and the sequences $s_\ell$ using wavelet matrices. The alphabet partitioning used is \texttt{dense} $\ell_{min} = \lg{23}$.
This was the most competitive combination for \texttt{AP} in our tests.

\item \texttt{BLCD-WT}: the original wavelet trees \cite{GGVsoda03} (\texttt{wt\_int} in \texttt{sdsl}), using plain bit vectors (\texttt{bit\_vector<>} in \texttt{sdsl}), and constant time $\rank$ and $\select$ using \texttt{rank\_support\_v} and \texttt{select\_support\_mcl}, respectively, from the \texttt{sdsl}.

\item \texttt{GMR}: the data structure by Golynski et al.~\cite{GMRsoda06} (\texttt{gmr\_wt<>} in
\texttt{sdsl}), using samplings 4, 8, 16, 32, and 64 for the inverse permutation data structure.

\item \texttt{HUFF-WT}: the Huffman-shaped wavelet trees, using plain bit vectors (\texttt{bit\_vector<>} in \texttt{sdsl}) for the WT nodes, with constant-time $\rank$ and $\select$ on them using classes \texttt{rank\_support\_v} and \texttt{select\_support\_mcl} from the \texttt{sdsl}. This approach achieves $H_0(s)$ compression thanks to the Huffman shape of the WT.

\item \texttt{RRR-WT}: the original wavelet trees \cite{GGVsoda03} (\texttt{wt\_int} in \texttt{sdsl}), this time using compressed bit vectors by Raman et al.~\cite{RRRtalg07} for the WT nodes (\texttt{rrr\_vector<>} in the \texttt{sdsl}, using block sizes 15, 31, and 63). This approach also achieves $H_0(s)$ compression, this time by compressing each WT node. 

\item \texttt{HUFF-RRR-WT}: the Huffman-shaped wavelet trees, using compressed bit vectors by Raman et al.~\cite{RRRtalg07} for the WT nodes. As in the previous approach, we used \texttt{rrr\_vector<>} in the \texttt{sdsl}, with block sizes 15, 31, and 63.

\item \texttt{WM}: the wavelet matrix \cite{CNPis15}, using a plain bit vector (\texttt{bit\_vector<>} in \texttt{sdsl}) and constant time $\rank$ and $\select$ using \texttt{rank\_support\_v} and \texttt{select\_support\_mcl} from the \texttt{sdsl}.
\end{itemize}

\begin{figure}[ht]
\centering
\includegraphics[width=0.49\textwidth]{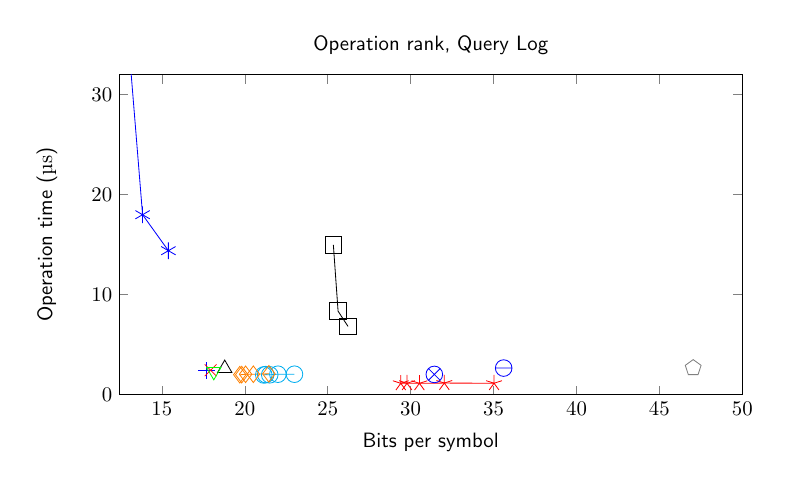}
\includegraphics[width=0.49\textwidth]{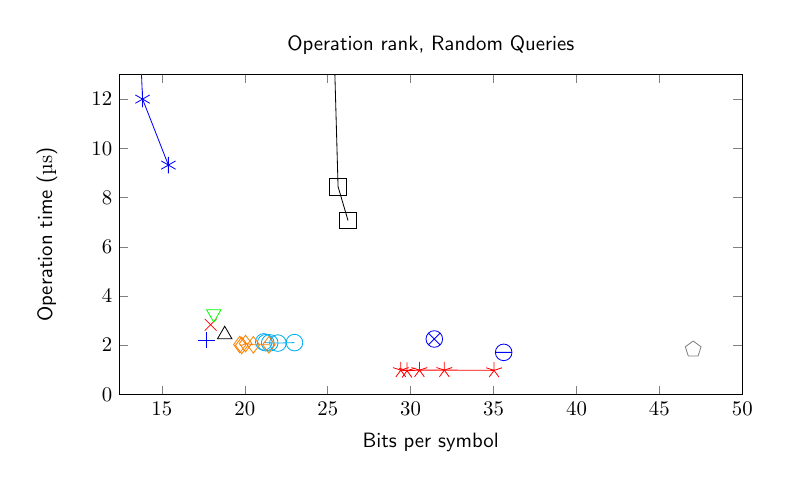}

\includegraphics[width=0.49\textwidth]{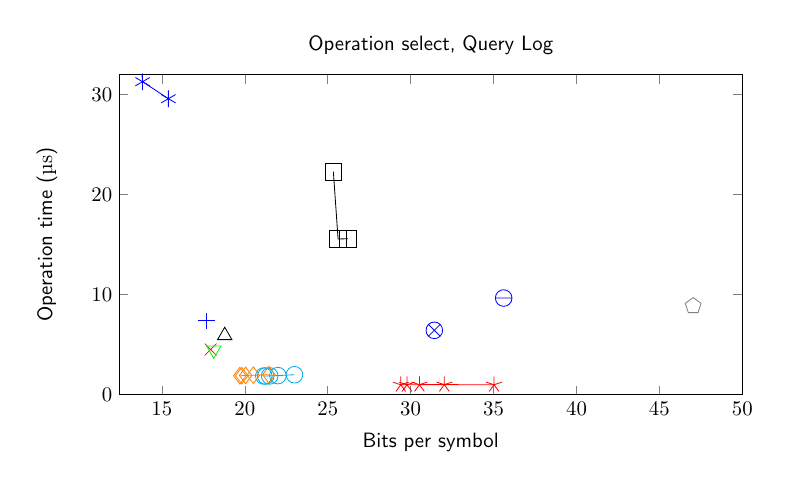}
\includegraphics[width=0.49\textwidth]{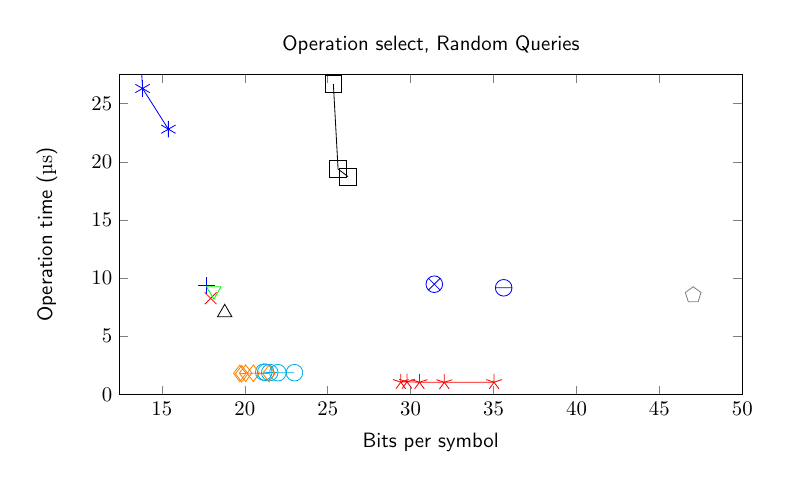}

\includegraphics[width=0.7\textwidth]{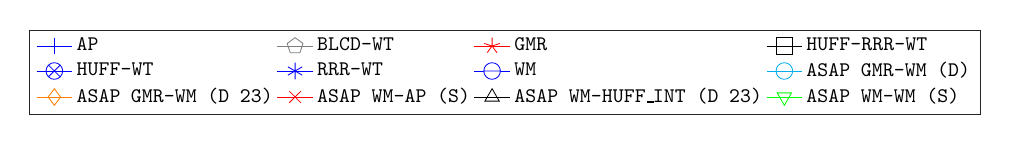}

\caption{Experimental results for operations $\rank$ (above) and $\select$ (below) on the Wikipedia text.
The $x$ axis shows the space usage in bits per symbol and starts at $H_0(s) = 12.45$ bits. The $y$ axis shows the average operation time, in microseconds per operation.}
\label{fig:rank-select}
\end{figure}

Figure \ref{fig:rank-select} shows the experimental results for operations
$\rank$ and $\select$, comparing the above approaches.
As it can be seen, \texttt{ASAP} yields interesting trade-offs.
In particular, for operation $\select$ and random symbols, alternative
\texttt{ASAP GMR-WM (D 23)} uses 1.11 times the space of \texttt{AP}, while improving the average time per $\select$, being 
4.85 times faster on average (from 9.37 to 1.92 microseconds per $\select$).
For query-log symbols, we obtain similar results.
However, this time there is another interesting alternative: \texttt{ASAP WM-AP (S)}
uses only 1.01 times the space of \texttt{AP} (i.e., an increase of about 1\%),
while being 1.63 times faster on average. 
For $\rank$ queries, \texttt{ASAP} is 
1.05--1.21 times faster on average.
In this case, the improvements are smaller compared to $\select$. This is because operation $\rank$ on bit vectors $\texttt{sd\_vector<>}$ 
is not as efficient as $\select$ \cite{OSalenex07}, whereas the original \texttt{AP} implements $\rank$ on the Huffman-shapped WT (for $t$) using the much faster $\texttt{rank\_support\_v}$ class in the \texttt{sdsl}.

\begin{figure}[ht]
\centering
\includegraphics[width=0.5\textwidth]{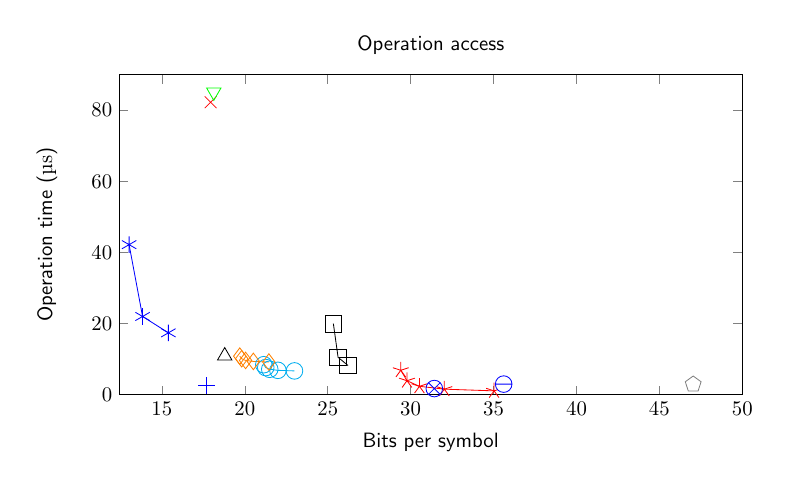}
\includegraphics[width=0.7\textwidth]{Plots/Basic-Operations/leyenda.pdf}
\caption{Experimental results for operation access.
The $x$ axis shows the space usage in bits per symbol and starts at $H_0(s) = 12.45$ bits. The $y$ axis shows the average operation $\access$ time, in microseconds per operation.}
\label{fig:access}
\end{figure}

Figure \ref{fig:access} shows experimental results for operation $\access$.
As expected, we cannot compete with the original \texttt{AP} scheme. However,
we are still faster than \texttt{RRR-WT}, and competitive with \texttt{GMR} \cite{GMRsoda06},
yet using much less space.

\section{Experimental Results on Information-Retrieval Applications}

After the promising results of our approach for supporting the fundamental string operations, 
we test in this section our alphabet-partitioning implementation on some relevant information-retrieval applications.

\subsection{Application 1: Snippet Extraction}

\begin{figure}[ht]
\centering
\includegraphics[width=0.49\textwidth]{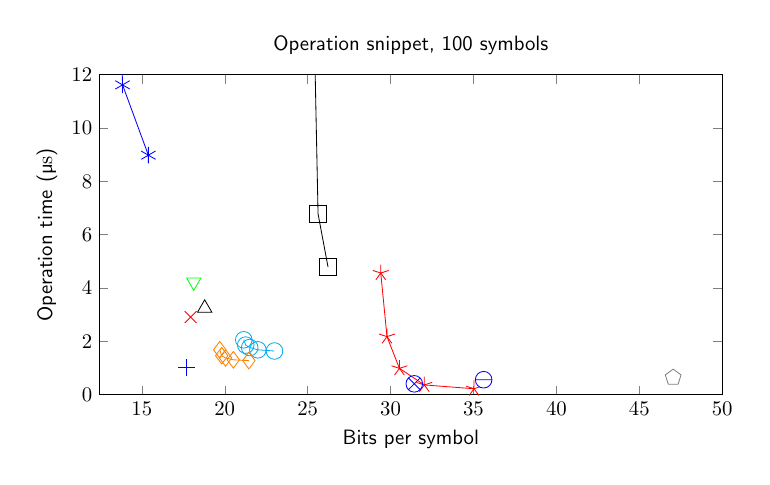}
\includegraphics[width=0.49\textwidth]{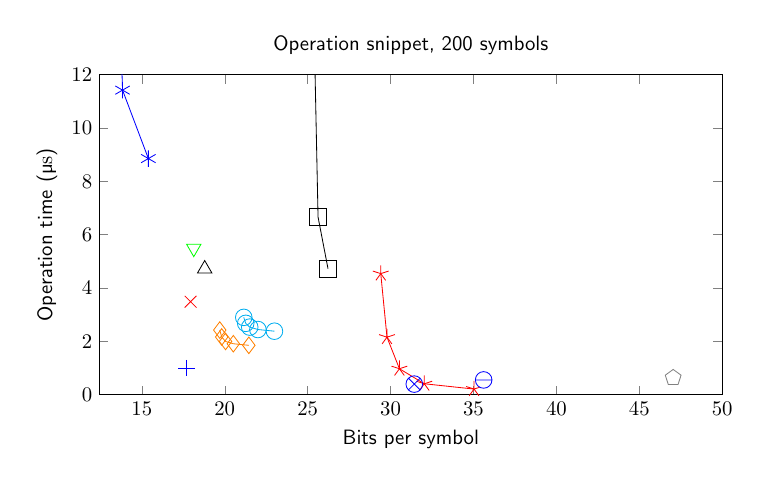}
\includegraphics[width=0.7\textwidth]{Plots/Basic-Operations/leyenda.pdf}

\caption{Experimental results for extracting snippets of length  $L=100$ (right) and $L=200$ (left).
The $x$ axis shows the space usage in bits per symbol and starts at $H_0(s) = 12.45$ bits. The $y$ axis shows the average extraction time, in microseconds per symbol extracted.}
\label{fig:snippets}
\end{figure}

We evaluate next the snippet extraction task, common in text search engines \cite{AGMOSVis20,TTHWsigir07}.
As we have already said, in this case one needs operation $\access$ to obtain not just a single symbol, but a snippet $s[i..i+L-1]$ of $L$ consecutive symbols in $s$.
In our experiments we tested with $L= 100$ and $L=200$  and generated 10{,}000 positions $i$ uniformly at random in the interval $[1{..}n-L]$. See Figure \ref{fig:snippets} for the experimental results. 
As it can be seen, we are able to reduce the time per symbol considerably 
(approximately 4 times faster ) when
compared with operation $\access$, making our approach
more competitive for snippet extraction.
It is important to note that
operation $B_j.\select$ in line 7 of Algorithm \ref{alg:snippet}
is implemented using the $\select$ operation provided
by the  $\texttt{sd\_vector<>}$ implementation.

\subsection{Application 2: Intersection of Inverted Lists}

Another relevant application of $\rank$/$\select$
data structures is that of intersecting inverted lists.
Previous work \cite{AGOspire10} has shown that one can simulate the intersection of inverted lists by representing the document collection (concatenated into a single string) with a $\rank$/$\select$ data structure. So, within the compressed space used by the text collection, one is able to simulate:
\begin{itemize}
\item the inverted index of the text collection, supporting intersection of (the simulated) inverted lists in time close to that of Barbay and Kenyon adaptive intersection time \cite{BKtalg08,AGOspire10};

\item a positional inverted index, using the approach by Arroyuelo et al.~\cite{AGMOSVis20}; and

\item snippet extraction.
\end{itemize}
These cover most functionalities of a search engine on document collections \cite{CB15}, making it an attractive approach because of its efficient space usage. An interesting application is that of allowing users of an IR system to carry out local searches. The idea is that the user can download the document collection (or a subset of it) using our compressed representation, thus supporting all the above functionalities using no additional space. 

\begin{figure}[ht]
\centering
\includegraphics[width=0.5\columnwidth]{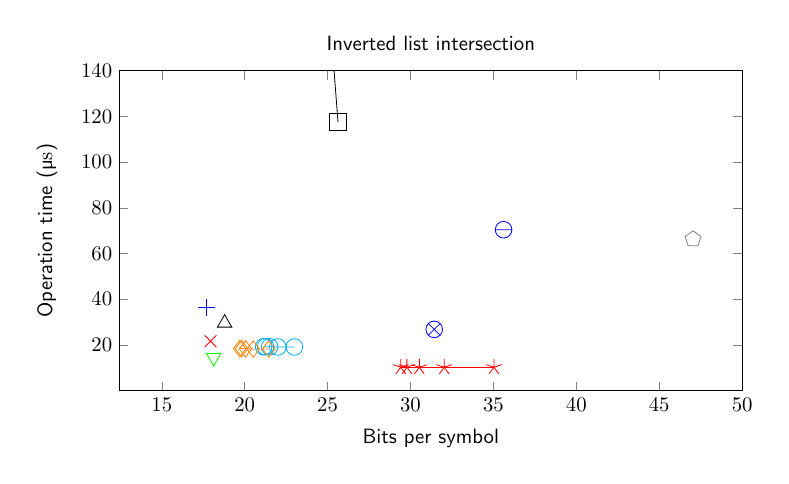}
\includegraphics[width=0.7\textwidth]{Plots/Basic-Operations/leyenda.pdf}
\caption{Experimental results for inverted list intersection. Times are in milliseconds.
The $x$ axis starts at $H_0(s) = 12.45$ bits.}
\label{fig:intersection}
\end{figure}

Figure \ref{fig:intersection} shows experimental results for intersecting
inverted lists. 
We implemented the variant of intersection algorithm tested by Barbay et al.~\cite{BCGNNalgo14}.
As it can be seen, \texttt{ASAP} yields important improvements in this application:
using only 2\% extra space, \texttt{ASAP wm-wm (S)} is able to improve the intersection time of
\texttt{AP}, being 2.54 times faster on average. 
The intersection times we observe ($\sim 14$ milliseconds for the most-competitive approach) are competitive with those of boolean AND queries using different variants of compressed inverted indexes \cite{YDSwww09,PVtois17,AOGSipm18,AGMOSVis20}.

\section{Alphabet Partitioning for Representing Strings with Runs} \label{sec:run-strings}


Let us consider next the case where the input string $s$ is formed by $r$ runs of equal symbols, $s = c_1^{l_1}c_2^{l_2}\cdots c_r^{l_r}$, where $c_i \in \Sigma$, for $i=1,\ldots, r$, $l_1, l_2, \ldots, l_r > 0$, $c_i \not = c_{i+1}$ for all $1\le i < r$, and $2\le \sigma \le r$. We shall show that alphabet partitioning introduces additional advantages when used to represent these strings.
We study first how to implement run-length compression with the original alphabet partitioning approach \cite{BCGNNalgo14}, to then analyze the scheme resulting from our approach.

\subsection{Partitioning into Subalphabets and Mapping $m$}
As a first observation, 
note that the alphabet partition does not necessarily need to be carried out according to Equation (\ref{eq:sparse}) nor, alternatively, the \texttt{dense} approach used in practice, whose main objective is to partition the original alphabet in such a way $H_0(s)$ compression is still achieved. As we aim at run-length compression now, we can partition in a different ---much simpler--- way that will allow us to prove interesting theoretical bounds.
The idea is to divide alphabet $\Sigma$ into $p = \lceil\lg^c{n}\rceil$ subalphabets consisting of $\lceil\sigma/p\rceil$ symbols each (alphabet $\Sigma_{p-1}$ can contain less symbols), for any integer constant $c \ge 1$. Now, for all $\alpha \in \Sigma$ we define:
\begin{equation}
m[\alpha] = \alpha\bmod{p}.
\end{equation}
Notice that $m[\alpha]$ can be computed on the fly when needed, so there is no need to store mapping $m$, saving space. A symbol $\alpha \in \Sigma$ assigned to subalphabet $\Sigma_\ell$ is reenumetared as $\lfloor \alpha / p \rfloor$ within $\Sigma_\ell$.

\subsection{Subalphabet Strings}

Another important fact that is worth noticing is that when representing a string with runs $s = c_1^{l_1}c_2^{l_2}\cdots c_r^{l_r}$ with the alphabet partitioning approach, these $r$ runs are not only retained by mapping $t$ and the sub-alphabet strings $s_{\ell}$, but also the lengths of the runs can be potentially increased (which, as a consequence, implies that the number of runs is decreased). That is, the number of runs in mapping $t$ and strings $s_{\ell}$ can be smaller than $r$, as we note next:
\begin{itemize}
    \item For $t$, if symbols $c_j, c_{j+1}, \ldots, c_{j+k}$ (for $1\le j \le r$, $k\ge 0$, and $j+k \le r$) correspond to the same sub-alphabet, then these $j+k+1$ runs of string $s$ become a single run of length $l_j+l_{j+1}+\cdots+l_{j+k}$ in $t$. Let us call $r^{(t)} \le r$ the resulting number of runs in mapping $t$. 

    \item Similarly, there can be symbols $c_j = c_{j+k} = \cdots = c_{j+q}$ (i.e., equal symbols whose runs are not consecutive in $s$) that could form a single run of length $l_j + l_{j+k} + \cdots + l_{j+q}$ in the corresponding string $s_{\ell}$. Let us call $r_{\ell}$, for $\ell=0,\ldots, p-1$, the resulting number of runs in $s_{\ell}$. Then, let us denote $r^{(s)} = \sum_{\ell=0}^{p-1}{r_{\ell}} \le r$. 
\end{itemize}
This potential for decreasing the overall number of runs is one of the advantages of using alphabet partitioning for run-length encoded strings, as we show next in this section and the experiments of Section \ref{sec:run-string-experiments}. 

\subsection{Our Representation}

For mapping $t$, which has alphabet of size $p=\lceil\lg^c{n}\rceil$ (for any integer constant $c \ge 1$), we use Fuentes-Sepúlveda et al.~\cite{FKKPdcc18} data structure, requiring $(1+\epsilon)r^{(t)}\lg{\left(\frac{n\lg^c{n}}{r^{(t)}}\right)} + \bigoh{r^{(t)}}
= (1+\epsilon)r^{(t)}\left(\lg{\left(\frac{n}{r^{(t)}} + c\lg\lg{n}\right)}\right) + \bigoh{r^{(t)}}
$ bits. For the sub-alphabet strings, we concatenate them into a single string $s'[1{..}n] = s_0\cdot s_1  \cdots s_p$ of length $n$ and up to $r^{(s)}$ runs \footnote{Notice that since the original symbols are re-enumerated within each sub-alphabet, there can be less than $r^{(s)}$ runs after concatenating all $s_\ell$, however we use $r^{(s)}$ as an upper bound in our analysis.}. The alphabet of $s'$ has size $\sigma_{\ell} = \lceil\sigma/\lceil\lg^c{n}\rceil\rceil$. We also use Fuentes-Sepúlveda et al.'s data structure for $s'$, requiring 
$(1+\epsilon)r^{(s)}\lg{\left(\frac{n\sigma}{r^{(s)}\lg^c{n}}\right)} + \bigoh{r^{(s)}}$ additional bits. Extra $\bigoh{\lg^c{n}}$ space is needed to delimit the ranges of $s'$ corresponding to each $s_\ell$. 
Operation $s.\rank$ is supported in
$\bigoh{
\lg{\frac{\lg{\frac{n\lg^c{n}}{r^{(t)}}}}{\lg\lg{n}}} +
\lg{\frac{\lg{\frac{n\sigma}{r^{(s)}\lg^c{n}}}}{\lg\lg{n}}}
}$ time, where the first term comes from computing $t.\rank$ and the second from $s_i.\rank$, for the corresponding subalphabet sequence $s_i$ (recall Equation (\ref{eq:rank-ap})). Operation $\select$ takes 
$\bigoh{\lg{\frac{\lg{(n/\min{\{r^{(t)},r^{(s)}\}})}}{\lg\lg{n}}}}$ time, where $\min{\{r^{(t)},r^{(s)}\}}$ comes from computing $t.\select$ and $s_i.\select$ (recall Equation (\ref{eq:select-ap})). Finally, operation $\access$ is supported in
$
\bigoh{
\lg{\frac{\lg{\frac{n\lg^c{n}}{r^{(t)}}}}{\lg\lg{n}}} +
\lg{\frac{\lg{(n/r^{(s)})}}{\lg\lg{n}}}
}
$ time, where the first term comes from computing $t.\rank$ and the second from $s_i.\select$ (recall Equation (\ref{eq:access-ap})).


Next, we consider replacing $t$ with bit vectors $B_0, \ldots, B_p$. In particular, we concatenate them into a single bit vector $B[1{..}n\lceil\lg^c{n}\rceil]$, with $n$ \bit{1}s and $r^{(t)}$ runs of $\bit{1}$s. According to Arroyuelo and Raman \cite{ARalgo22}, bit vector $B$ can be represented using $\lg{n\lg^c{n} - n + 1 \choose r^{(t)}} + \lg{n-1 \choose r^{(t)}-1} + o(n)$ bits, which by Stirling approximation is about $r^{(t)}\lg{\left(\frac{n\lg^c{n}}{r^{(t)}}\right)} + r^{(t)}\lg{\left(\frac{n}{r^{(t)}}\right)} + o(n) = 2r^{(t)}\lg{\left(\frac{n}{r^{(t)}}\right)} + r^{(t)}\cdot c\lg{\lg{n}} + o(n)$ bits.
Operation $\rank$ is supported in time $\bigoh{\lg{\frac{\lg{(n\sigma/(r^{(s)}\lg^c{n}))}}{\lg\lg{n}}}}$, whereas $\select$ takes time $\bigoh{\lg{\frac{\lg{(n/r^{(s)})}}{\lg\lg{n}}}}$. Operation $\access$ takes $\bigoh{\lg^c{n} + \lg{\frac{\lg{(n/(r^{(s)}\lg^c{n}))}}{\lg\lg{n}}}}$. Notice that the running time of operations $\rank$ and $\select$ are not bigger than the ones corresponding to the scheme described in the previous paragraph. This could yield improved performance in practice. On the other hand, this scheme uses slightly more space (depending on the value of $\epsilon$) than the space used by $t$ of the original alphabet-partitioning approach.
However, we claim that by using bit vector $B$ instead of $t$ allows for a faster implementation in practice, as we shall evidence in our experiments of Section \ref{sec:run-string-experiments}. 
We summarize the performance of our resulting data structures in Table \ref{tab:wt-runs-state-of-the-art} (on page \pageref{tab:wt-runs-state-of-the-art}).



\section{Application 3: Full-Text Search on large-Alphabet Highly-Repetitive Collections} \label{sec:run-string-experiments}

%


Next, we address the classical full-text search problem, defined as follows. Let $T[1{..}n]$ be a text string of length $n$ over an alphabet $\Sigma  = \{0, \ldots, \sigma-1\}$. 
Let $P[1{..}M]$ be another string (the search pattern) of length $M$ over the same alphabet $\Sigma$. The problem consists in finding (or counting) all occurrences of $P$ in $T$. If the text is given in advance to queries and several queries will be issued on it, one can afford building a data structure on the text to later speed-up query processing time. Several solutions exist for this classical problem, such as \emph{suffix trees} \cite{Weiner73,McCreight76,ACFGMcacm16}, \emph{suffix arrays} \cite{MMsicomp93}, and \emph{compressed full-text indexes} \cite{FMjacm05,GGVsoda03,ANSalgo12}. We show next that our alphabet-partitioning implementation for strings with runs from Section \ref{sec:run-strings} is a competitive building block for implementing \emph{compressed-suffix arrays} such as FM-indexes \cite{FMjacm05,GNPjacm20}. We particularly focus on highly-repetitive text databases, which have flourished in several scenarios such as (to name just a few): 
\begin{enumerate}\item  \emph{Biological databases}, where the DNA of different individuals of the same species share a high percentage of their sequences \cite{MBCT2015}. These databases need to be searched for patterns that are of interest to biologists; 

\item \emph{Source-code repositories}, where the different (highly-similar) versions of a source code are stored such that users can query them; and 

\item \emph{The Wikipedia Project}, where the different (highly-similar, in general) versions of Wikipedia pages need to be stored and searched \cite{HSsigir12}. 
\end{enumerate}
We focus on large-alphabet texts, as these are the most suitable for the alphabet-partitioning approach. 

FM-indexes build on the Burrows-Wheeler transform (BWT, for short) of text $T$. The BWT is an elegant approach used for compression boosting \cite{BWT} while providing indexed-search functionalities on the compressed text \cite{FMjacm05}. Given a text $T[1{..}n]$ such that $T[n] = `\$'$ is a special text terminator that does not occur in $T$, the main idea is to (conceptually) build a table $\mathcal{B}[1{..}n][1{..}n]$ such that row $i$ of the table corresponds to the circular shift $T[i{..}n]\cdot T[1{..}i-1]$ of the text. Then, the rows of this table are sorted lexicographically. After this sorting step, let $F[1{..}n]$ and $L[1{..}n]$ denote the first and last column of table $\mathcal{B}$, respectively, both of which are needed to support efficient text searching. As the rows of $\mathcal{B}$ are lexicographically sorted, column $F$ is a sorted array that starts with a run corresponding to symbol $0$, then a run of symbol $1$, and so on, with $\sigma$ runs overall. The length of each such run equals the number of occurrences of the corresponding symbol in $T$. Hence, $F$ can be implemented efficiently using an array $C[0{..}\sigma-1]$ such that $C[i]$ is the total number of symbols in $T$ that are smaller than symbol $i$. Array $C$ alone is enough to represent array $F$, using $\sigma \lg{n}$ bits of space. 

Column $L$, on the other hand, is actually the Burrows-Wheeler transform of $T$. The backward search algorithm by Ferragina and Manzini \cite{FMjacm05} allows one to count the number of occurrences of a pattern $P$ in $T$ using the BWT of $T$. We only need to represent string $L$ using a data structure for $L.\rank$. Algorithm \ref{alg:backward-search} illustrates the process, which carries out at most $2M$ $\rank$ operations on the BWT $L$ (see lines \ref{line:5} and \ref{line:6}).  

\begin{algorithm}[ht]
\centering
\caption{$\mathsf{BackwardSearch}(P[1{..}M])$}
\label{alg:backward-search}
\begin{algorithmic}[1]
\STATE $b \gets 1$
\STATE $e \gets n$
\FOR{$i \gets M$ \textbf{downto} $1$} 
\STATE $c \gets P[i]$
\STATE $b \gets C[c] + L.\rank_c (b-1)+1$ \label{line:5}
\STATE $e \gets C[c] + L.\rank_c (e)$ \label{line:6}
\IF{$b > e$}{\STATE \textbf{break}}
\ENDIF
\ENDFOR
\RETURN $e-b$
\end{algorithmic}
\end{algorithm}

For highly-repetitive texts, column $L$ tends to have a few long runs of equal symbols \cite{GNPjacm20}, which can be exploited to improve compression. Although the process explained above to build $L$ might suggest that $\Theta(n^2)$ space and time is needed (in order to build table $\mathcal{B}$), it can be carried out in $\Theta(n)$ time and space using the suffix array of $T$ \cite{Navarro2016}. Let $r$ denote the number of runs in the BWT of a text $T$. RLFM-indexes \cite{MNnjc05} are one of the most effective approaches for counting the number of occurrences of a pattern in $T$. They use $\bigoh{r}$ words of space and support the count operation in $\bigoh{M\lg\lg_w{(\sigma + \frac{n}{r})}}$ time \cite{GNPjacm20}.

Next, we test this application in practice. We represent the BWT of text $T$ using our \texttt{ASAP} data structure and implement Algorithm \ref{alg:backward-search} on it. 
In our experiments, we just count the occurrences of $P$ in $T$, disregarding the process of actually locating the text positions of the occurrences as these need to store additional information that could dominate the space usage and blur our main results. Also, on suffix-array based indexes, the process of finding the occurrences succeeds the counting process. So, a faster count implies an earlier start of the locating step. 
Our experimental setup is the same as in Section \ref{sec:experimental-setup}. Our source code is available at
\url{https://github.com/carlosrojasmo/ASAP-RLFMindex}. See also \url{https://github.com/Yhatoh/ASAP_paper/}, which includes all the source code produced in this paper.

\subsection{Text Collections}
We test with the BWT of the following highly-repetitive texts from the \emph{Pizza\&Chili Corpus} \cite{FGNVjea08} \footnote{\url{http://pizzachili.dcc.uchile.cl/repcorpus.html}.}:
\begin{itemize}
\item \textbf{Einstein.de}: A single sequence containing the concatenation of the different versions of the Wikipedia articles corresponding to Albert Einstein, in German. The original text can be obtained from \url{http://pizzachili.dcc.uchile.cl/repcorpus/real/einstein.de.txt.gz}. We enumerate all words and punctuation marks in the text and use a 32-bit integer array to represent it. The result is a sequence of 29{,}970{,}946 integers, with an alphabet of size 7{,}105 \footnote{In our experiments, the alphabet size is the number of different integer word ids we generate}. The BWT of this text has 47{,}422 runs. We also tested the English.en text available from \url{http://pizzachili.dcc.uchile.cl/repcorpus/real/einstein.en.txt.gz}, with similar results.


\item \textbf{English}: The sequence obtained from the concatenation of English text files selected from \texttt{etext02} to \texttt{etext05} collections of the Gutenberg Project. This text can be obtained from
\url{http://pizzachili.dcc.uchile.cl/repcorpus/pseudo-real/english.001.2.gz}. We use the same representation as before (using 32-bit integers to represent words and punctuation marks), obtaining a text of 39{,}894{,}587 integers and alphabet of size 89{,}753. The number of runs in the BWT of this text is 682{,}453.

\item \textbf{Coreutils}: 9 versions of the Coreutils 5.x  package, that can be downloaded from
\url{http://pizzachili.dcc.uchile.cl/repcorpus/real/coreutils.gz}.
The resulting sequence has 93{,}319{,}681 integers and alphabet of size 148{,}654. The number of runs in the BWT of this text is 2{,}723{,}451, which is larger than the other texts, indicating that this text could be less compressible.


\end{itemize}
We also tested with other repetitive text available at Pizza\&Chili Corpus, namely Einstein.en (the English version of Einstein.de) and the world-leaders text, obtaining similar results to the above ones. 
Table \ref{tab:repetitive-texts} shows a summary of the main statistics of the tested texts. 
\begin{table}[ht]
    \centering       
    \caption{Statistics of the highly-repetitive texts used in our tests.}
    \label{tab:repetitive-texts}

\medskip

    \begin{tabular}{lcccc}
    \toprule
Text & Length & $\sigma$ & BWT runs & Avg.~R.~L. \\
    \midrule
     Einstein.de & 29{,}970{,}946 & 7{,}105 & 47{,}422 & 632   \\
     English     & 39{,}894{,}587 & 89{,}753 & 682{,}453 & 58  \\
     Coreutils   & 93{,}319{,}681 & 148{,}654 & 2{,}723{,}451 & 34\\
     \bottomrule
\end{tabular}

\end{table}

\subsection{Practical Partition Approaches} \label{sec:ap-rl-partitions}

We test the following four partition approaches:
\begin{itemize}
\item{A1:} We divide the alphabet into $p=\lceil\lg{n}\rceil$ subalphabets consisting of $\lceil\sigma/\lceil\lg{n}\rceil\rceil$ alphabet symbols each (the last partition can have less symbols). In this way,
\begin{equation}
m[\alpha] = \left\lfloor\frac{\alpha}{\lceil\sigma/\lceil\lg{n}\rceil\rceil}\right\rfloor.
\end{equation}
Notice that $m[\alpha]$ can be computed on the fly when needed, so there is no need to store mapping $m$. This could save non-negligible space \cite{BCGNNalgo14}. A symbol $\alpha \in \Sigma$ assigned to subalphabet $\Sigma_{\ell}$ is re-enumerated as $\alpha \bmod{(\lceil\sigma/\lceil\lg{n}\rceil\rceil)}$ within $\Sigma_{\ell}$. 

\item{A2:} We divide the alphabet again into $p=\lceil\lg{n}\rceil$ subalphabets consisting of $\lceil\sigma/\lceil\lg{n}\rceil\rceil$ alphabet symbols each (the last partition can have less symbols). Now, we define
\begin{equation}
m[\alpha] = \alpha \bmod \lceil\lg{n}\rceil.
\end{equation}
As for the previous alternative, $m[\alpha]$ can be computed on the fly when needed, hence there is no need to store mapping $m$. Every symbol $\alpha$ corresponding to subalphabet $\Sigma_{\ell}$ is re-enumerated as $\lfloor \alpha / \lceil \lg{n}\rceil\rfloor$ within $\Sigma_{\ell}$.

\item{A3:} The \texttt{dense} partitioning \cite{BCGNNalgo14} that we already mentioned in Section \ref{sec:AP}, such that
\begin{equation}
    m[\alpha] = \lfloor\lg{r(\alpha)} \rfloor,
\end{equation}
where $r(\alpha)$ denotes the ranking of symbol $\alpha$ according to its frequency.

\item{A4:} A variant of the \texttt{dense} approach from above, but the symbol ranking is carried out according to the string $c_1c_2\cdots c_r$, i.e., the run headers. 
\end{itemize}

Note that for alternatives A1 and A2, $p = \lceil\lg{n}\rceil$, whereas for A3 and A4 it holds that $p = \lceil\lg{\sigma}\rceil$. Alternative partition approaches could be defined, yet the ones we define are enough for our purposes: theoretical and practical efficiency. Next, we summarize their main features:
\begin{itemize}
\item
Approaches A1 and A2 do not store mapping $m$, saving space. More importantly, A1 and A2 do not need to compute $m.\access$ (to obtain the subalphabet corresponding to a given symbols) and $m.\rank$ (to translate a given symbol into the corresponding symbol within its subalphabet). Hence, the overall $s.\rank_\alpha$ and $s.\select_\alpha$ processing time could be improved in practice. 
On the other hand, as a consequence of the way approaches A1 and A2 distribute the symbols within the subalphabets, it holds that $|\Sigma_{p-1}| \le |\Sigma_{0}| = \cdots = |\Sigma_{p-2}| = \lceil\sigma/\lceil\lg{n}\rceil \rceil$. Therefore, $s_i.\rank$ and $s_i.\select$ are carried out on subalphabets of size $\lceil\sigma/\lceil\lg{n}\rceil\rceil$, which can worsen performance (when compared to A3 and A4, see below).

\item Approaches A3 and A4, on the other hand, have subalphabets whose size increases exponentially among partitions. Assuming $\sigma = 2^k$, for $k\ge 1$, the subalphabet sizes are $\sigma/2^i$, for $1\le i \le \lg{\sigma}$. So, notice that $\sigma / 2^i \le \sigma/\lg{n}$ whenever $i \ge \lg\lg{n}$ holds. This means that out of the $\lg{\sigma}$ partitions of A3 and A4, only $\lg\lg{n}$ of them have alphabet bigger than the corresponding ones in A1 and A2. In the remaining $\lg{\sigma} - \lg\lg{n}$ partitions, operations $s_i.\rank$ and $s_i\select$ have the potential to be computed more efficiently in practice (because of their smaller alphabet). On the other hand, A3 and A4 need to compute operations $m.\access$ and $m.\rank$, increasing the overall computation time. 
\end{itemize}

Table \ref{tab:repetitive-texts-partition-approaches} shows the experimental values for $r^{(t)}$ and $r^(s)$ for each text and partition approach we tested. These values must be compared with the number of runs in the BWT of each text, which is shown in the upper table. 
\begin{table}[ht]
    \centering       
    \caption{Statistics for the different alphabet partition alternatives we propose.}
    \label{tab:repetitive-texts-partition-approaches}

\medskip
    \begin{tabular}{clccc}
    \toprule
Partition  &  Text  & $r^{(t)}$ & $r^{(s)}$ & $p$  \\
    \midrule
A1 &     Einstein.de  &  25{,}502 & 39{,}033 & 13  \\
   &     English      &  348{,}464 & 581{,}470 & 13  \\
   &     Coreutils    &  1{,}063{,}354 & 2{,}388{,}078 & 14 \\
    \midrule
A2 &     Einstein.de  &  44{,}267 & 25{,}305 & 13  \\
   &     English      &  635{,}291 & 417{,}815 & 13  \\
   &     Coreutils    &  2{,}561{,}930 & 1{,}354{,}863 & 14 \\
    \midrule
A3 &     Einstein.de  &  10{,}650 & 44{,}260 & 13   \\
   &     English      &  185{,}017 & 629{,}518 & 17   \\
   &     Coreutils    &  436{,}282 & 2{,}649{,}860 & 18 \\
    \midrule
A4 &     Einstein.de  &  7{,}428 & 46{,}610 & 13    \\
   &     English      &  165{,}530 & 645{,}700 & 17   \\
   &     Coreutils    &  373{,}859 & 2{,}679{,}462 & 18  \\
    \bottomrule
    \end{tabular}
\end{table}
For A1 and A2, we found out that $p = \lceil\lg{(n)/2} \rceil$ yields better results for our texts. So, for Einstein.de we have $p=13$, hence $\lceil\sigma/p\rceil = 546$ is the size of the subalphabet of each partition. For the English text we have $p=13$ and $\lceil\sigma/p\rceil = 6{,}904$, and for Coreutils $p=13$ and $\lceil\sigma/p\rceil = 10{,}618$.  

\subsection{Practical Run-Length Compressed Bit Vectors}

In this and the next subsection we implement the main components of the data structure from Section \ref{sec:run-strings}.
We survey in this section approaches for compressing bit vectors with runs of \bit{1}s, in order to decide the best alternative for $B_0, \ldots, B_{p-1}$. We complement the study by Boffa et al.~\cite{BFVtalg22} by testing a new scenario (the BWT of highly-compressible texts) as well as new approaches for representing bit vectors with runs.
In particular, we test the following approaches.
\begin{itemize}
        \item $\mathtt{PEF}$: The \emph{Partitioned Elias-Fano} (PEF) approach \cite{OVsigir14}. We re-implemented this scheme based on the original one, adding the support for $\rank$ and $\select$ (the original one only supported the next-greater-or-equal operation). We only test with the Optimized Partition approach (\texttt{opt}), as this is the one that uses the least space. The main idea is that the original bit vector is divided into blocks such that: (1) each block is represented either as a plain bit vector, an \texttt{SDArray}, or implicitly if it consists only of \bit{1}s; and
        (2) the overall space usage obtained from this partitioning is optimized \footnote{Or almost \cite{OVsigir14}, as an approximation algorithm is used to compute the partitioning.}.

        We parameterized the minimum size a block can have as 64, 128, 256, 512, 1024, and 2048. So, $\mathtt{PEF}$ determines the optimal block sizes such that no block can be smaller than these values. For the blocks represented with plain bit vectors, we tested with \texttt{bit\_vector<>} from the \texttt{sdsl} and Vigna's broad-word approach \cite{Vwea08} (\url{https://github.com/vigna/sux}). For the former, we tested with \texttt{rank\_support\_v} and \texttt{rank\_support\_v5} in order to provide different space/time trade-offs. However, and since we use these approaches to represent relatively small bit vectors, we modified the original implementation of both rank support data structures such that the precomputed rank information now is stored using an array of $\lg{u}$-bit integers, being $u$ the length of the bit vector. The original implementation used an array of 64-bit integers. We divided the bit vector into blocks of 512 bits. Hence, we store a second-level array of 9-bit integers, as the precomputed values in this level are local to a block. For blocks represented with \texttt{SDArray} we use \texttt{sd\_vector<>} from the \texttt{sdsl}.
        The source code of our implementation can be accessed from our code repository \cite{our-code}. 
        
        \item $\mathtt{S18}$: The $\rank$/$\select$ data structure based on S18 compression \cite{AOGSipm18,AWsccc20}. We used block sizes  8, 16, 32, 64, 128,  and 256. The source code of this approach can be also accessed from our repository \cite{our-code}. 
        
        \item $\mathtt{HYB}$: The \texttt{hyb\_vector<>} data structure \cite{KKPdcc14} from the \texttt{sdsl}, using block size 16. 
        
        \item $\mathtt{Zombit}$: The data structure by Gómez-Brandón \cite{Gomez-Brandondcc20}. This data structure only implements operation $\rank$, and the source code can be accessed at \url{https://github.com/adriangbrandon/zombit/blob/main/include/zombit_vector.hpp}.
        
        \item $\mathtt{OZ}$: The approach by Delpratt et al.~\cite{DRRwea06}, which stores the lengths of the \bit{0} and \bit{1} runs using two separate arrays, \texttt{Z} and \texttt{O} \cite[see Section 4.4.3]{Navarro2016}. We use an implementation provided by Gómez-Brandón \cite{Gomez-Brandondcc20}, where these arrays are represented using \texttt{sd\_vector<>} from the \texttt{sdsl}.
        
        \item $\mathtt{RLE\_VECTOR}$: The approach tested by Boffa et al.~\cite{BFVtalg22}, which implements scheme \texttt{OZ} yet using VNibble compression \cite{AGMOSVis20}.

        \item $\texttt{LA}$: The \emph{learned approach} by Boffa et al.~\cite{BFVtalg22}, using parameters 1, 2, 3, 6, 10, 15, and 18. The version $\texttt{LA-OPT}$ \cite{BFVtalg22} corresponds to parameter 1 (which is the one using the least space, as well as the fastest).
\end{itemize}
Additionally, we include the basic compressed approaches from the \texttt{sdsl}, namely $\texttt{sd\_vector<>}$ \cite{OSalenex07}, which we call $\mathtt{SD}$ in our plots, and $\texttt{rrr\_vector<>}$ \cite{RRRtalg07} (using blocks of size 15, 31, 63, and 127), which we call $\mathtt{RRR}$.

\begin{figure}
    \centering
    \hbox{\includegraphics[width=0.49\textwidth]{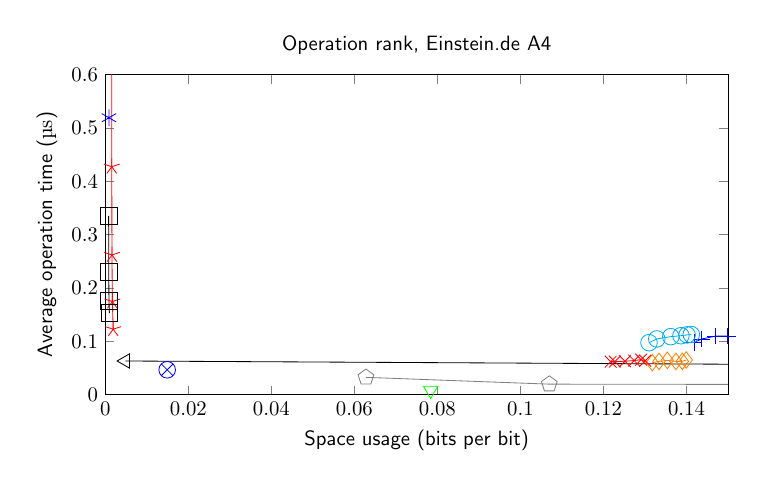}
    \includegraphics[width=0.49\textwidth]{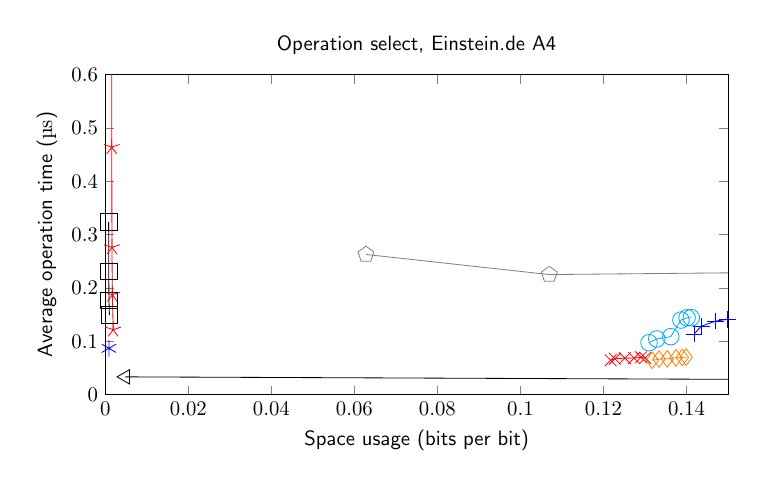}}
    \hbox{
    \includegraphics[width=0.49\textwidth]{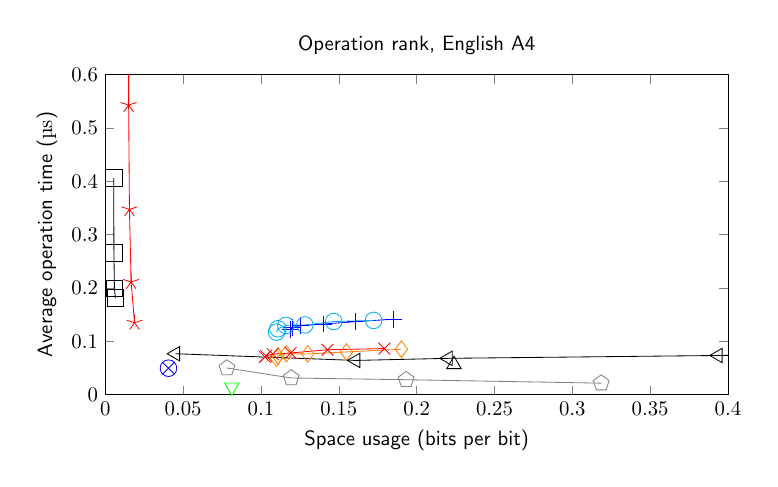}
    \includegraphics[width=0.49\textwidth]{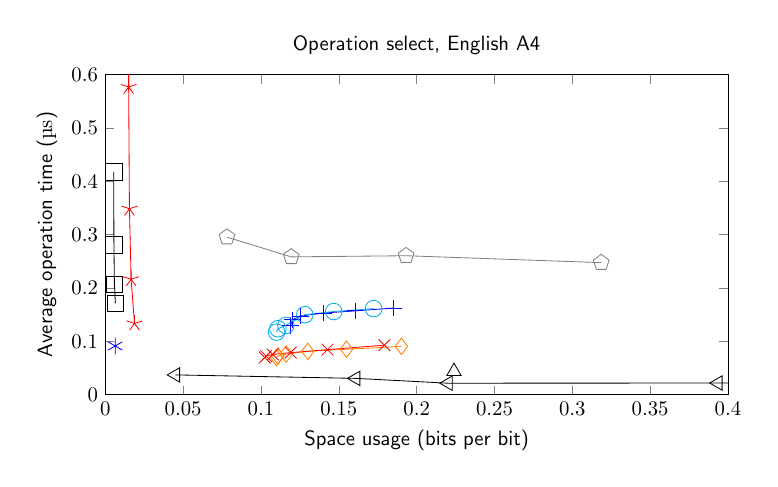}}
    \hbox{
    \includegraphics[width=0.49\textwidth]{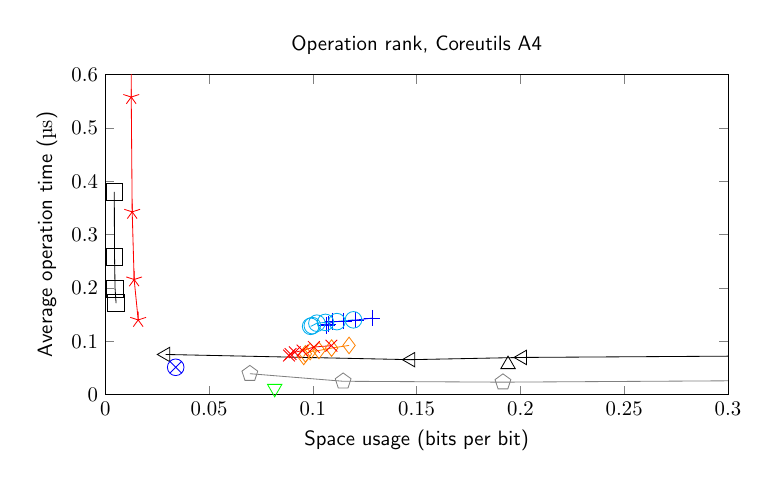}
    \includegraphics[width=0.49\textwidth]{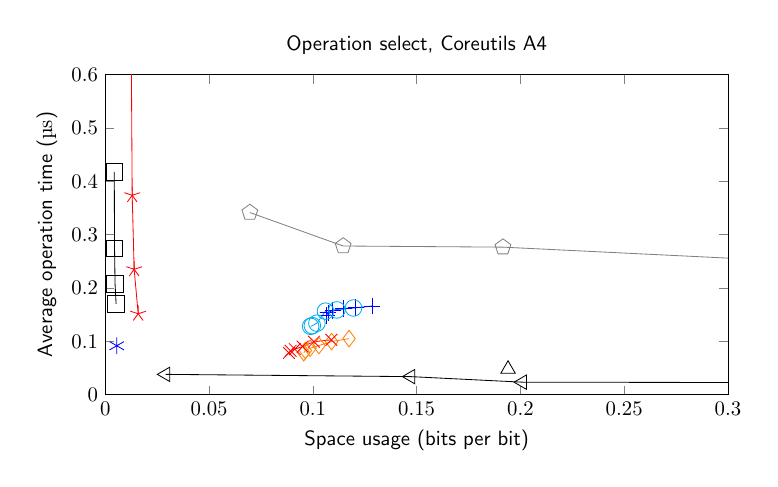}}
    \includegraphics[width=0.7\textwidth]{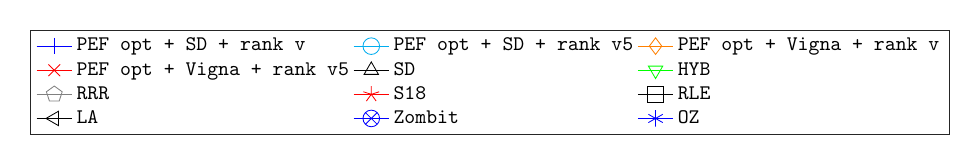}
    \caption{Experimental space/time trade-offs for different compressed bit vector representations on bit sequences with runs. The bit vectors used in the tests correspond to sequences $B_0,\ldots, B_{p-1}$ using partition approach A4. Results for operation $\rank$ are shown in the left column, whereas $\select$ is shown in the right column.}
    \label{fig:rank-select-B-runs}
\end{figure}

In order to choose the most efficient trade-offs for our scenario, we test experimentally using 1{,}000{,}000 random $\rank$ and $\select$ queries on the $p$ bit vectors of the ASAP data structure corresponding to each text we tested (the parameters for $\rank$ and $\select$ and the bit vector on which the query is carried out, are generated uniformly at random). Figure \ref{fig:rank-select-B-runs} shows the space-time trade-off for operations $\rank$ and $\select$ on the bit vectors $B_0, \ldots, B_{p-1}$ corresponding to partition approach A4 (similar conclusions are obtained for the remaining partition schemes). Regarding space usage, we report the average over the $p$ bit vectors for each text.
As it can be seen, \texttt{RLE\_VECTOR} and \texttt{S18} offer the most competitive trade-offs for $\rank$ and $\select$, so we will use them as building block for our approach in the experiments of Section \ref{sec:experiments-RL-strings}. \texttt{OZ} is also competitive, yet just for $\select$. As backward search only needs $\rank$, we disregard it in what follows. 

These results also support one of our main claims: that using $B_0, \ldots, B_{p-1}$ to implement $t$ yields better performance in practice. As it can be seen, $\rank$ is supported in about 100--200 nanoseconds by the most competitive data structures in all tested datasets---there are data structures supporting $\rank$ in less than 100 nanoseconds, yet they have a higher space usage. For $\select$, a similar conclusion can be achieved. On the other hand, operation $t.\rank$ would be carried out on some kind of wavelet tree by the original alphabet partitioning approach, which means several $\rank$ operations on the bit vectors that represent the wavelet tree nodes, needing more time.

\subsection{Practical Run-Length Compressed Strings} \label{sec:sub-alphabet-seqs-practice}
For the sub-alphabet strings $s_0,\ldots, s_{p-1}$ we test the following run-length-compression approaches:
\begin{itemize}
    \item \texttt{RLMN}: the run-length wavelet tree by M{\"{a}}kinen and Navarro \cite{MNnjc05}, implemented with class \texttt{wt\_rlmn<>} in the \texttt{sdsl}. 
    
    \item \texttt{FKKP}: the run-length data structure by Fuentes-Sepúlveda et al.~\cite{FKKPdcc18}. We use the original implementations by the authors, obtained from the \texttt{sdsl} fork by Fuentes-Sepúlveda \cite{fkkp-code}.
\end{itemize}
We tried the possible combinations among these approaches and \texttt{RLE\_VECTOR} or \texttt{S18}, obtaining the 8 schemes denoted by the following regular expression (with `$|$' meaning `or'): 
\[
\texttt{ASAP}~\underbrace{(\texttt{RLMN(AP)}~|~\texttt{RLMN(INT)}~|~\texttt{FKKP(AP)}~|~\texttt{FKKP(GMR)})}_{\textrm{representation~for~} s_0,\ldots, s_{p-1}}~~\underbrace{(\texttt{RLE}~|~\texttt{S18})}_{\textrm{representation~for~} B_0, \ldots, B_{p-1}}
\]
The alternatives for the second term in the regular expression (corresponding to the representation of  $s_0, \ldots, s_{p-1}$) are as follows:
\begin{itemize}
    
    \item \texttt{RLMN(AP)} and \texttt{RLMN(INT)}: the approach by M\"akinen and Navarro \cite{MNnjc05}, using the \texttt{wt\_rlmn<>} implementation from the \texttt{sdsl}. We use \texttt{AP} as the base data structure (first approach) and a wavelet tree  \texttt{wt\_int<>} from the \texttt{sdsl} (second approach).

    \item \texttt{FKKP(AP)} and \texttt{FKKP(GMR)}: the approach by Fuentes-Sep\'ulveda et al.~\cite{FKKPdcc18}, using \texttt{AP} \cite{BCGNNalgo14} and \texttt{GMR} \cite{GMRsoda06} (\texttt{wt\_gmr<>} in the \texttt{sdsl}) as building blocks. 
\end{itemize}

\subsection{Experimental Results}
\label{sec:experiments-RL-strings}

In our experiments, we searched for 50{,}000 unique random patterns of length $M = 4, 8$, and $16$ words. To generate the patterns, we choose positions $i$ uniformly at random in the interval $[1{..}n-M]$ and then use $T[i{..}i+M-1]$ as a pattern. This ensures that there is at least one occurrence of each pattern in the text.

We compared with the following baseline approaches, which we shall use to represent the original BWT of each text:
\begin{itemize}
    \item \texttt{AP(RLMN)}: the alphabet partitioning approach \cite{BCGNNalgo14} implemented by class \texttt{wt\_ap<>} in the \texttt{sdsl}, using the \texttt{RLMN} approach \cite{MNnjc05} for $t$ and the sub-alphabet sequences $s_i$, corresponding to class \texttt{wt\_rlmn<>} from the \texttt{sdsl}).
    \item \texttt{RLMN(AP)}: a particular implementation of the so-called RLFM-index \cite{MNnjc05}, on which the r-index \cite{GNPjacm20} is built. The setup is similar to that of Section \ref{sec:sub-alphabet-seqs-practice}
;
    \item \texttt{FKKP(AP)} and \texttt{FKKP(GMR)}: using the same setup as described in Section \ref{sec:sub-alphabet-seqs-practice} for the subalphabet sequences. 
\end{itemize}

\begin{figure}
    \centering
    \includegraphics[width=0.49\textwidth]{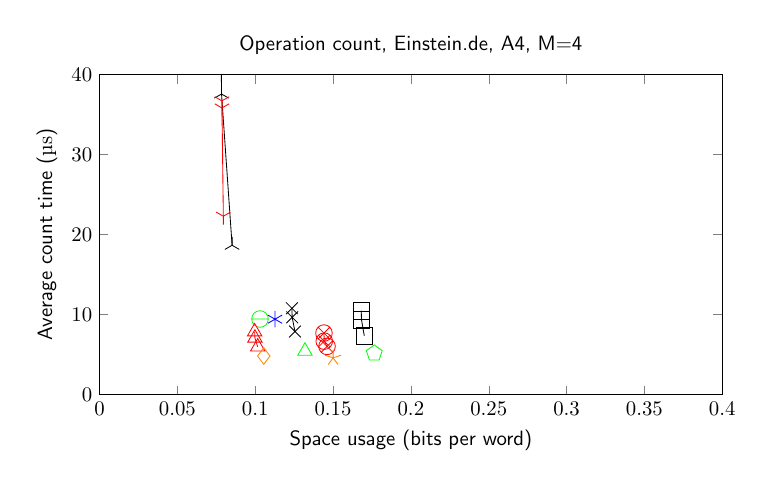}
    \includegraphics[width=0.49\textwidth]{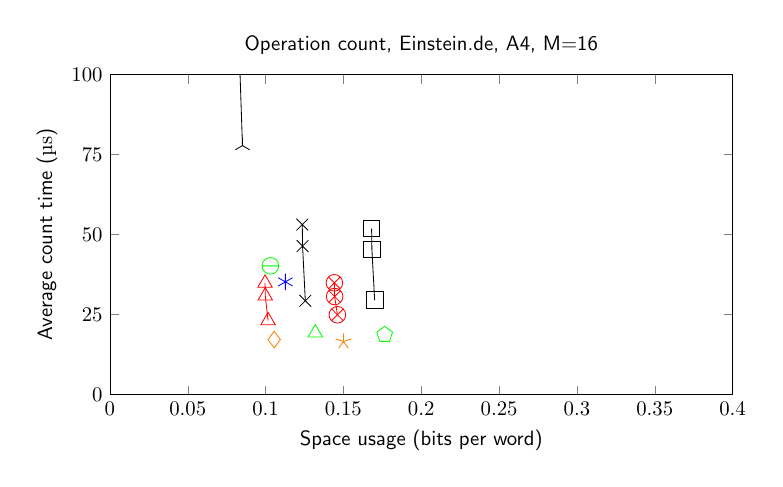}

    \includegraphics[width=0.49\textwidth]{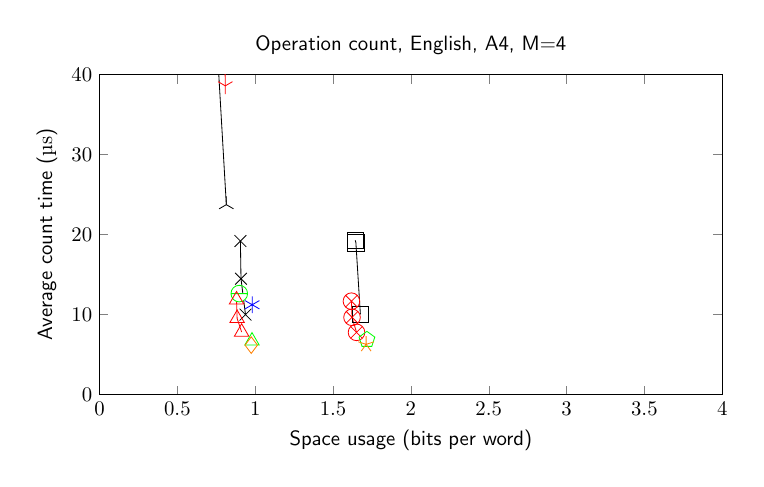}
    \includegraphics[width=0.49\textwidth]{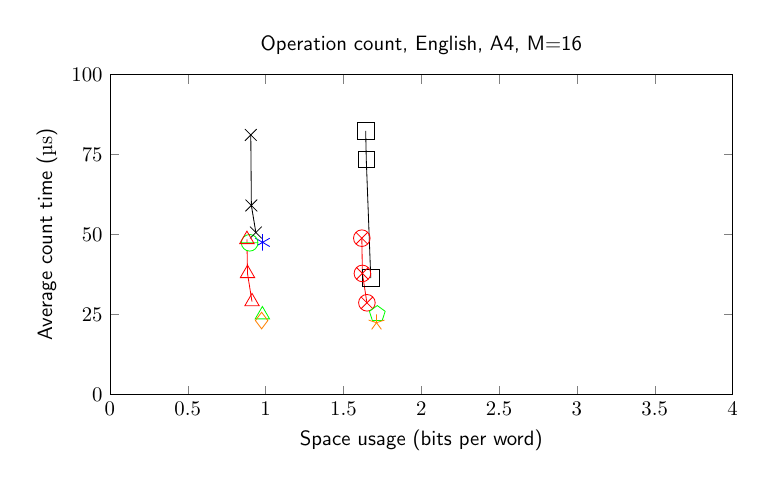}
    \includegraphics[width=0.49\textwidth]{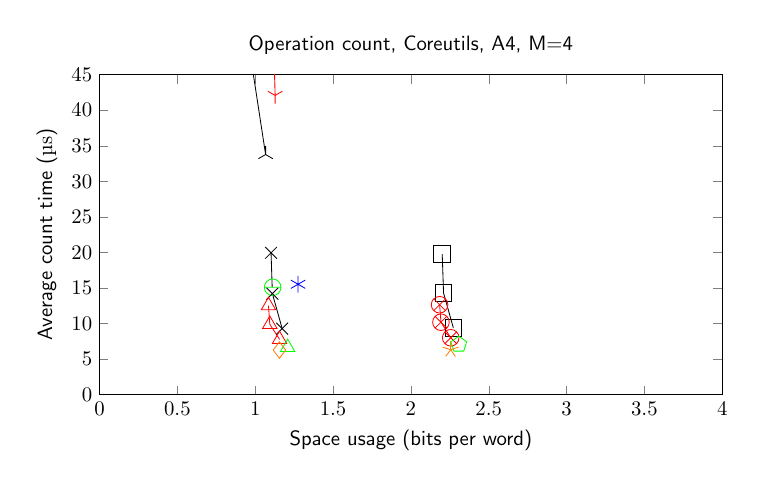}
    \includegraphics[width=0.49\textwidth]{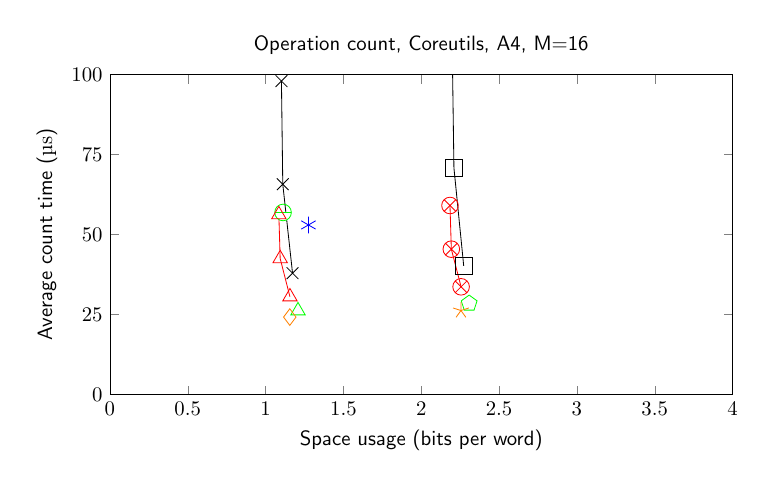}
     \includegraphics[width=0.7\textwidth]{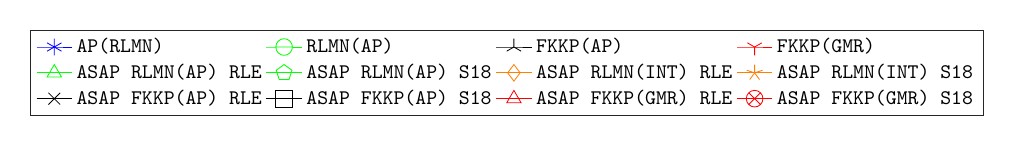}

    \caption{Space/time trade-off for count queries on the Burrows-Wheeler Transform (see Algorithm \ref{alg:backward-search}) of different large-alphabet texts, for the alphabet partition alternative A4 using patterns of length 4 and 16.}
    \label{fig:count-BWT-A4}
\end{figure}

Figure \ref{fig:count-BWT-A4} 
shows the experimental space/time trade-offs for counting the number of occurrences of a pattern using Algorithm \ref{alg:backward-search} \cite{FMjacm05}, for partition approach A4 (which is slightly more efficient than A3, and considerably better than A1 and A2). Experimental results for A1, A2, and A3 are shown in \ref{sec:additional-experiments},  Figures \ref{fig:count-BWT-A1}, \ref{fig:count-BWT-A2}, and \ref{fig:count-BWT-A3}, respectively. We show the average search time per pattern searched, and the space usage of each scheme, in bits per text word.

In all cases, schemes \texttt{RLMN(AP)} 
is the most competitive one among the state-of-the-art approaches, so we will use it as baseline for comparison. In particular, \texttt{RLMN(AP)} uses less space than \texttt{AP(RLMN)} in all cases, with comparable query performance. Regarding \texttt{FKKP}, it turns out to be the most space-efficient among all alternatives (in some cases being considerably smaller), yet at the price of a less competitive query-processing time.

%
%
%
%

\subsubsection{Comparison Between A2 and A4}

The main question we raised in Section \ref{sec:ap-rl-partitions} is whether A1 or A2 yield more efficient representations than A3 or A4, as both have advantages and drawbacks. 
According to our experiments, A2 and A4 are the most efficient representatives on each side, so we shall compare them. We use \texttt{ASAP RLMN(INT) RLE} for the comparison, being the best performer. 

Table \ref{tab:ASAP-components-runs} shows the space usage of the main components of \texttt{ASAP RLMN(INT) RLE}, as well as the space usage of mapping $t$ (to compare with the space usage of bit vectors $B_0, \ldots, B_{p-1}$).
From these results we can conclude that mapping $m$ uses negligible space compared to the remaining components. 
\begin{table}[ht]
    \centering       
    \caption{Space usage (in bits per text word) of the different components of the alphabet-partitioning data structures, for scheme \texttt{ASAP RLMN(INT) RLE}. We also show the size of mapping $t$, in order to compare with the space of bit vectors $B_0,\ldots, B_{p-1}$.}
    \label{tab:ASAP-components-runs}
\medskip
    \begin{tabular}{clccccc}
    \toprule
Partition  &  Text  & $t$ & $B_0,\ldots,B_{p-1}$ & $s_0,\ldots, s_{p-1}$ & $m$  \\
    \midrule
A1 &     Einstein.de  & 0.032 & 0,029 & 0.067 & --   \\
   &     English      & 0.236 & 0.202 & 0.667 & --  \\
   &     Coreutils    & 0.292 & 0.219 & 1.038 & --\\
    \midrule
A2 &     Einstein.de  & 0.054 & 0.050 & 0.045 & --  \\
   &     English      & 0.398 & 0.389 & 0.548 & --   \\
   &     Coreutils    & 0.620 & 0.550 & 0.675 & --  \\
    \midrule
A3 &     Einstein.de  & 0.012 & 0.051 & 0.036 & 0.003 \\
   &     English      & 0.144 & 0.414 & 0.405 & 0.012 \\
   &     Coreutils    & 0.141 & 0.535 & 0.511 & 0.007 \\
    \midrule
A4 &     Einstein.de  & 0.009 & 0.051 & 0.037 & 0.003 \\
   &     English      & 0.131 & 0.413 & 0.407 & 0.012 \\
   &     Coreutils    & 0.124 & 0.533 & 0.524 & 0.007  \\
    \bottomrule
    \end{tabular}
\end{table}
Notice that A2 generates slightly smaller $B_0,\ldots, B_{p-1}$ than A4. On the other hand, A4 yields noticeably smaller subalphabet sequences $s_0,\ldots,s_{p-1}$: 17\% smaller for Einstein.de, 26\% for English, and 22\% for Coreutils. As a result, A2 uses more space than A4: 
1.04 (Einstein.de), 1.11  (English), and 1.14 (Coreutils) times the space of \texttt{ASAP RLMN(INT) RLE} on A4.

Regarding query processing time, A4 is faster than A2. Query processing time of scheme \texttt{ASAP RLMN(INT) RLE} on A4 is: for Einstein.de, 1.33 ($M=4$) and 1.49 ($M=16$) times faster than the same scheme on A2; for English, 1.31 ($M=4$) and 1.52 ($M=16$) times faster than on A2; and for Coreutils, 1.44 ($M=4$) and 1.63 ($M=16$) times slower than on A4. This indicates that operation $s_i.\rank$ on potentially smaller-alphabet strings (i.e., A3 and A4) yields better overall query performance, even though the process involves  operations $m.\rank$ and $m.\access$.

\subsubsection{Results for Partition Approach A4.}

For partition approach A4, \texttt{ASAP RLMN(INT) RLE} offers, in general, the best trade-off among our schemes.  Its space usage is of 0.11 bits per word for the Einstein.de text, 0.98 bits per word for English, and 1.17 bits per word for Coreutils. These are remarkable compression ratios, which indicate the high repetitiveness of the texts, as well as the effectiveness of the approaches we use.

\begin{itemize}
    \item \textbf{Einstein.de Text:} In this case, \texttt{ASAP RLMN(INT) RLE} uses 1.03 times the space of \texttt{RLMN(MN)}, while being 1.96 ($M=4$) and 2.30 ($M=16$) times faster.
    \item \textbf{English Text:} In this case, \texttt{ASAP RLMN(INT) RLE} uses 1.09 times the space \texttt{RLMN(MN)}, while being 1.92 ($M=4$) and 2.05 ($M=16$) times faster. \item \textbf{Coreutils Text:} In this case, \texttt{ASAP RLMN(INT) RLE} uses 1.05 times the space of \texttt{RLMN(MN)}, while being 2.33 ($M=4$) and 2.27 ($M=16$) times faster.
\end{itemize}

Interesting trade-offs are also introduced by $\texttt{ASAP FKKP(GMR) RLE}$:
\begin{itemize}
    \item \textbf{Einstein.de Text:} In this case, \texttt{ASAP FKKP(GMR) RLE} uses 0.99 times the space of \texttt{RLMN(MN)}, while being 1.53 ($M=4$) and 1.72 ($M=16$) times faster.
    \item \textbf{English Text:} In this case, \texttt{ASAP FKKP(GMR) RLE} uses 0.98 times the space \texttt{RLMN(MN)}, while being 1.30 ($M=4$) and 1.23 ($M=16$) times faster. 
    \item \textbf{Coreutils Text:} In this case, \texttt{ASAP FKKP(GMR) RLE} uses 0.98 times the space of \texttt{RLMN(MN)}, while being 1.50 ($M=4$) and 1.30 ($M=16$) times faster.
\end{itemize}

\section{Distributed Computation of $\rank$ and $\select$}

The partitions generated by the alphabet partitioning approach are amenable for the distributed computation of
batches of $\rank$ and $\select$ operations.
In this section we explore ways to implement then on a distributed scheme.
In a distributed query processing system, there exists a specialized node that
is in charge of receiving the query requests (this is called the \emph{broker}) \cite{CB15}, 
and then distributes the computation
among a set of computation nodes (or simply \emph{processors}). The latter are in charge of carrying out the actual 
computation. Next, we study how the original alphabet partitioning scheme (\texttt{AP})
and our proposal (\texttt{ASAP}) can be implemented on a distributed system, allowing
the fast computation of batches of $\rank$ and $\select$ queries.

\subsection{A Distributed Query-Processing System Based on \texttt{AP}}
The sub-alphabet sequences $s_\ell$ are distributed among the computation nodes, 
hence we have $p$ processors in the system.
We also have a specialized broker, storing mappings $m$ and $t$. This is a drawback
of this approach, as these mappings become a bottleneck for the distributed
computation.

\subsection{A Distributed Query-Processing System Based on \texttt{ASAP}}
In this case, 
the sub-alphabet sequences $s_\ell$ and the bit vectors
$B_\ell$ are distributed among the computation nodes.
Unlike  \texttt{AP}, now each computation node acts as a broker:
we only need to replicate the mapping $m$ in them.
The overall space usage for this is $O(p\sigma\lg{\lg{p}})$
if we use an uncompressed WT for $m$. This is only
$O(\sigma\lg\lg{p}) = o(n)H_0(s)$ bits per processor \cite{BCGNNalgo14}.
In this simple way, we avoid having a specialized broker, but distribute
the broker task among the computation nodes. This avoids bottlenecks at the broker,
and can make the system more fault tolerant.

Queries arrive at a computation node, which uses mapping $m$ to
distribute it to the corresponding computation node.
For operation $s.\access(i)$, we carry out a broadcast operation, in order
to determine for which processor $\ell$, $B_\ell[i] =\mathtt{1}$; this is the one
that must answer the query.
For extracting snippets, on the other hand, we also broadcast the operation to all
processors, which collaborate to construct the desired snippet using the
symbols stored at each partition.

\subsection{Comparison}
The main drawback of the scheme based on \texttt{AP} is that it needs a specialized broker
for $m$ and $t$. Thus, the computation on these mappings is not distributed, lowering
the performance of the system.
The scheme based on \texttt{ASAP}, on the other hand, allows a better distribution: 
we only need to replicate mapping $m$ in each processor, with a small space penalty in practice.
To achieve a similar distribution with \texttt{AP}, we would need to replicate $m$ and $t$ in each processor,
increasing the space usage considerably. Thus, given a fixed amount of main memory for the system, the scheme based on \texttt{ASAP}
would be likely able to represent a bigger string than \texttt{AP}.

Table \ref{table:distributed} shows experimental results on a simulation of these schemes. We only consider
computation time, disregarding communication time. 
\begin{table}[ht]
\caption{Experimental results for the distributed computation of operations on a string.
Times are in microseconds per operation, on average (for extracting snippets,
it is microseconds per symbol). For $\rank$
and $\select$, the symbols used are from our query log.
Scheme \texttt{ASAP} implements the sequences $s_\ell$ using wavelet matrices,
whereas mapping $m$ is implemented using a Huffman-shaped WT. 
The partitioning is \texttt{dense} $\ell_{min} = \lg{23}$. The number
of partitions generated (i.e., computation nodes in the distributed system) is 46.}
\label{table:distributed}
\centering
\begin{tabular}{lcccccccccc}
\toprule
Operation & ~ & \multicolumn{3}{c}{\texttt{ASAP}} & ~ & \multicolumn{3}{c}{\texttt{AP}} & ~ & \texttt{AP}/\texttt{ASAP}\\
\cline{3-5} \cline{7-9} 
                         & & Time  & ~ & Speedup & & Time  & ~ & Speedup \\
\midrule
$\rank$          & & 0,373 &   &  8.03   & & 1.310 &   & 1.91    & & 3.51\\ 
$\select$        & & 0.706 &   &  8.41   & & 2.970 &   & 2.55    & & 4.21\\
$\access$        & & 1.390 &   &  8.11   & & 2.130 &   & 1.45    & & 1.53\\
$\mathsf{snippet}$       & & 0.466 &   &  6.96   & & 0.939 &   & 1.25    & & 2.02\\
\bottomrule
\end{tabular}
\end{table}
As it can be seen, \texttt{ASAP} uses the distributed system in a better way.
The average time per operation for operations $\rank$ and $\select$ 
are improved by about 71\% and 76\%, respectively, when compared with \texttt{AP}.
For extracting snippets, the time per symbol extracted is reduced by about 50\%.
Although the speedup for 46 nodes might seem not too impressive (around
7--8), it is important to note that our experiments are just a proof of concept.
For instance, the symbols could be distributed in such a way that
the load balance is improved.

\section{Conclusions}

Our alphabet-partitioning $\rank$/$\select$ data structure
offers interesting trade-offs. Using slightly more space than the original
alphabet-partitioning data structure from \cite{BCGNNalgo14} (e.g., an 11\% increase in espace usage), we are able
to improve the time for operation $\select$ by about $80\%$ (a speed-up of 4.85 on average).
The performance for $\rank$ can be improved between
4\% and 17\% (1.05--1.21).
For the inverted-list intersection problem, we showed improvements of about 60\% 
for query processing time (a speed-up of 2.54), using only 2\% extra space when compared to the original
alphabet-partitioning data structure. This makes this kind of data structures
more attractive for this relevant application in information retrieval tasks.

Regarding strings formed by equal-symbol runs, we showed that alphabet partitioning (and in particular our implementation) can improve the overall performance (i.e., space usage and query processing time) by:
\begin{itemize}
    \item Potentially decreasing the number of runs that need to be represented by the data structure, hence improving compression;
    \item Introducing an integer parameter $c \ge 1$ to Fuentes-Sepúlveda et al.'s data structure \cite{FKKPdcc18} that allows us to improve their performance. 
    \item Replacing mapping $t$ of the original alphabet partitioning data structure \cite{BCGNNalgo14} by corresponding run-length compressed bit vectors, which allow us to improve theoretical and practical performance (see, e.g., Table \ref{tab:wt-runs-state-of-the-art} and our experimental results from Section \ref{sec:experiments-RL-strings}).    
\end{itemize}
Overall, our experimental results on the particular scenario of representing compressed suffix arrays based on the Burrows-Wheeler transform indicate that for the running time of the count operation (which counts the number of occurrences of a query pattern in a text) we obtain a speed-up of 1.23-2.33  (compared to RLFM-indexes \cite{MNnjc05,GNPjacm20}), using only 0.98--1.09 times their space. 

It remains open the problem of designing better strategies for partitioning the alphabet such that the resulting $r^{(t)}$ and $r^{(s)}$ are further decreased, hence improving space usage.

We also studied how the alphabet-partitioning data structures can be used
for the distributed computation of $\rank$, $\select$, $\access$, 
and $\mathsf{snippet}$ operations.
As far as we know, this is the first study about the support of
these operation on a distributed-computing environment.
In our experiments, we obtained speedups from 6.96 to 8.41, for 46 processors.
This compared to 1.25--2.55 for the original alphabet-partitioning data structure.
Our results were obtained simulating the distributed computation,
hence considering only computation time (and disregarding communication time).
The good performance observed in the experiments allows us to think about a real
distributed implementation. This is left for future work, as well as a more
in-depth study that includes aspects like load balance, among others.

Overall, from our study we can conclude that our approach can be used to implement the ideas from Arroyuelo at al.~\cite{AGMOSVis20}, such that by representing a document collection as a single string using our data structure (hence using compressed space), one can: (i) simulate an inverted index for the document collection; (ii) simulate tf-idf information (using operation $\rank$); (iii) simulate a positional inverted index (using operation $\select$); (iv) to carry out snippet extraction to obtain snippets of interest from the documents. Implementing an information retrieval system based on these ideas is left as future work.

\section*{Acknowledgements}

Diego Arroyuelo was funded by ANID – Millennium Science Initiative Program – Code ICN$17\_002$, Chile. Part of this work was carried out when Diego Arroyuelo and Gabriel Carmona were at the Department of Informatics, Universidad T\'ecnica Federico Santa Mar\'ia, Chile. All the experiments in this paper were carried out on a server from that Department. Special thanks to Claudio Lobos and José Herrera for their support.

\bibliographystyle{plain}
\bibliography{ref}

\clearpage

\appendix

\section{Additional Experiments} 
\label{sec:additional-experiments}

\begin{figure}[h]
    \centering
    \includegraphics[width=0.4\textwidth]{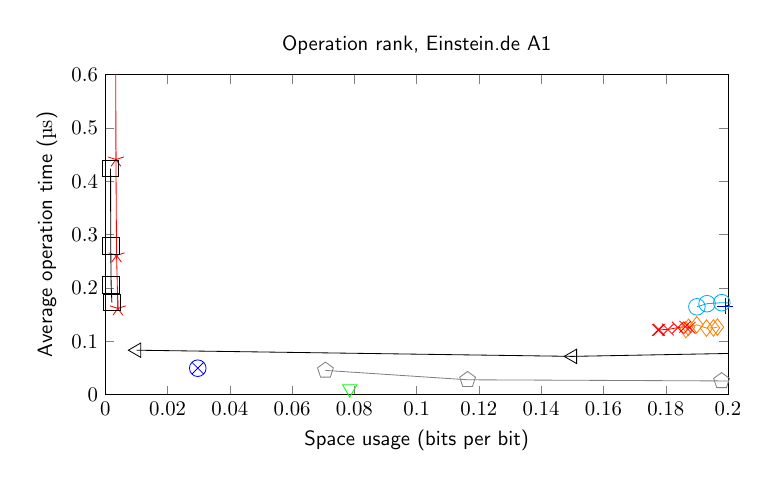}
    \includegraphics[width=0.4\textwidth]{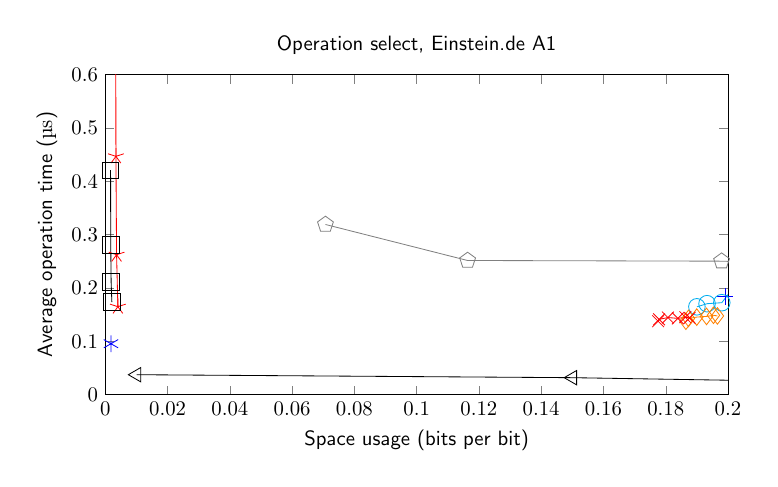}

    \includegraphics[width=0.4\textwidth]{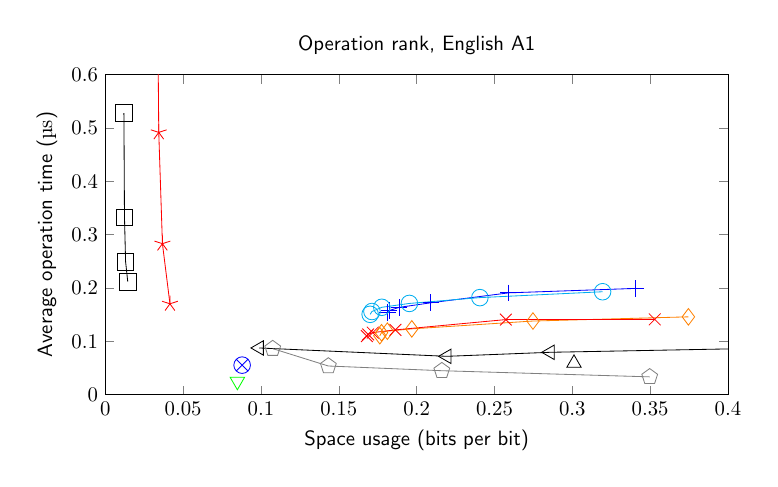}
    \includegraphics[width=0.4\textwidth]{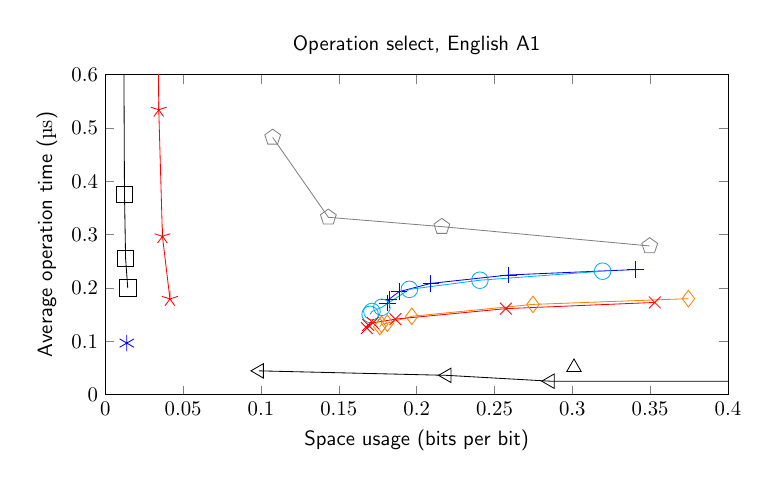}
    
    \includegraphics[width=0.4\textwidth]{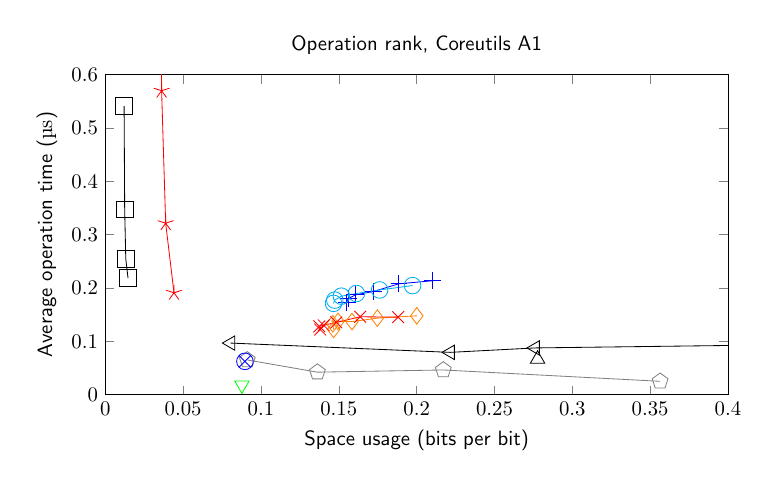}
    \includegraphics[width=0.4\textwidth]{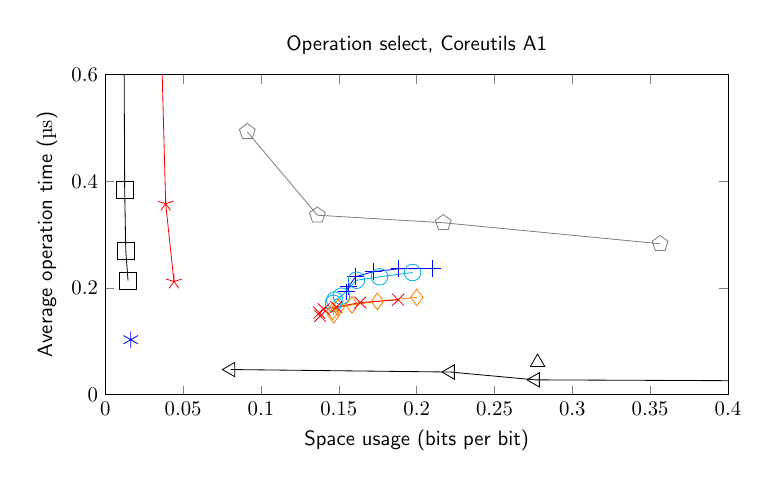}
    \includegraphics[width=0.7\textwidth]{Plots/BV2.0/leyenda_rank.pdf}
    \caption{Experimental space/time trade-offs for different compressed bit vector representations on bit sequences with runs. The bit vectors used in the tests correspond to sequences $B_0,\ldots, B_{p-1}$ using partition approach A1. Results for operation $\rank$ are shown in the left column, whereas $\select$ is shown in the right column.}
    \label{fig:rank-select-B-runs-A1}
\end{figure}

\begin{figure}[h]
    \centering
    \includegraphics[width=0.4\textwidth]{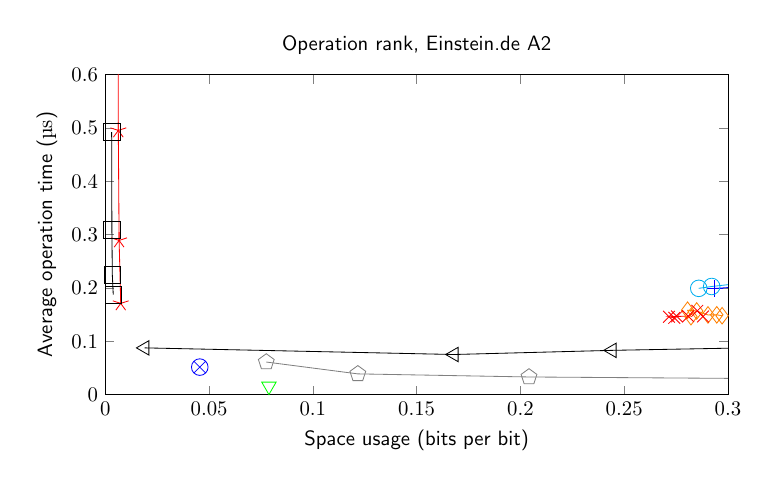}
    \includegraphics[width=0.4\textwidth]{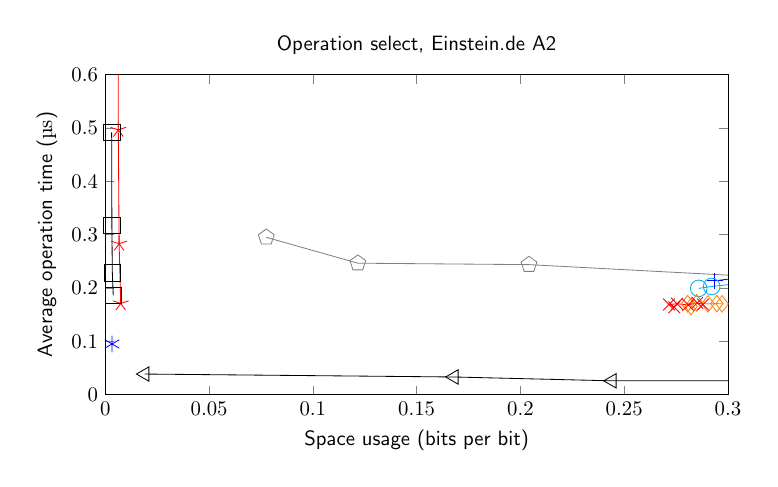}
    
    \includegraphics[width=0.4\textwidth]{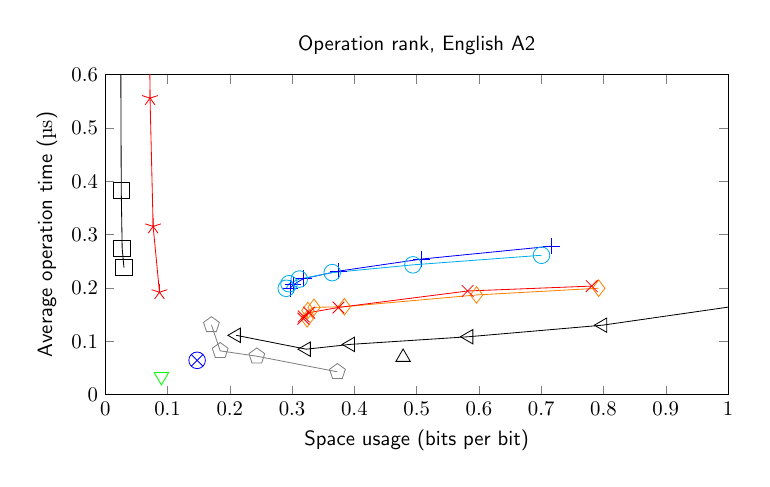}
    \includegraphics[width=0.4\textwidth]{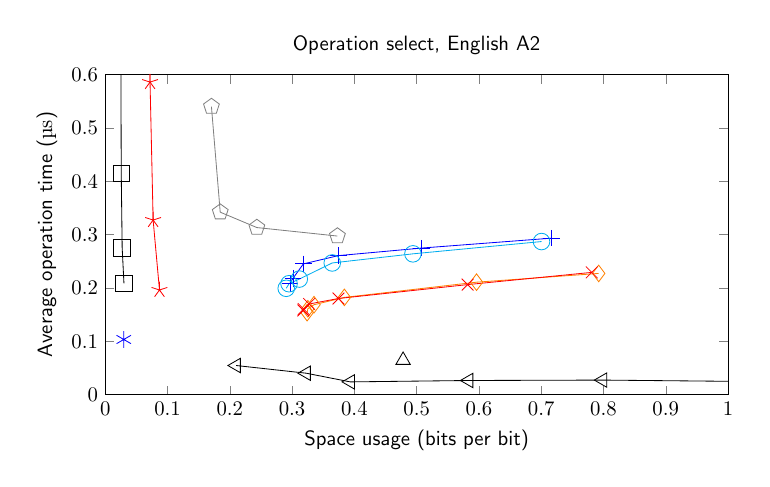}
    
    \includegraphics[width=0.4\textwidth]{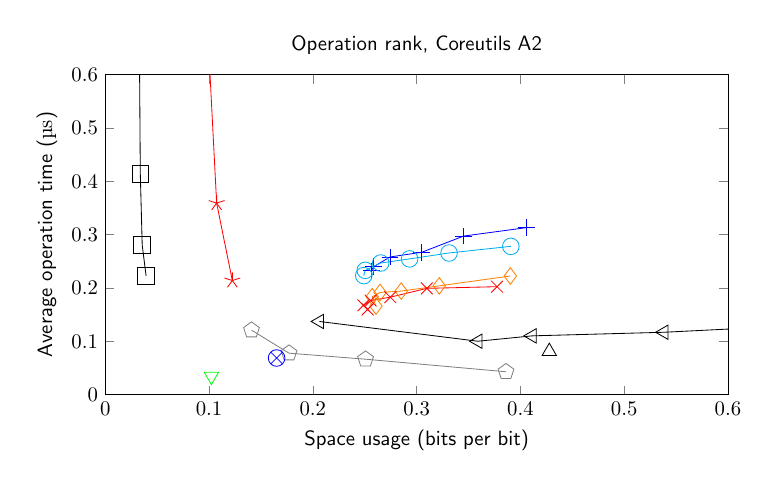}
    \includegraphics[width=0.4\textwidth]{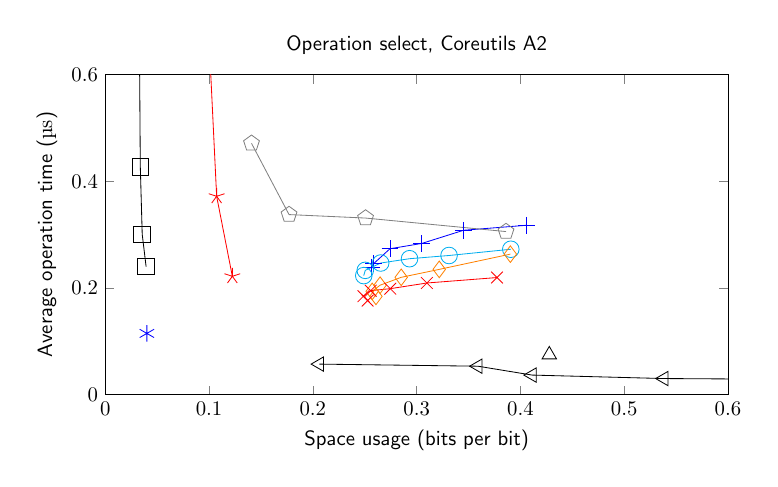}
    \includegraphics[width=0.7\textwidth]{Plots/BV2.0/leyenda_rank.pdf}
    \caption{Experimental space/time trade-offs for different compressed bit vector representations on bit sequences with runs. The bit vectors used in the tests correspond to sequences $B_0,\ldots, B_{p-1}$ using partition approach A2. Results for operation $\rank$ are shown in the left column, whereas $\select$ is shown in the right column.}
    \label{fig:rank-select-B-runs-A2}
\end{figure}

\begin{figure}[h]
    \centering
    \includegraphics[width=0.4\textwidth]{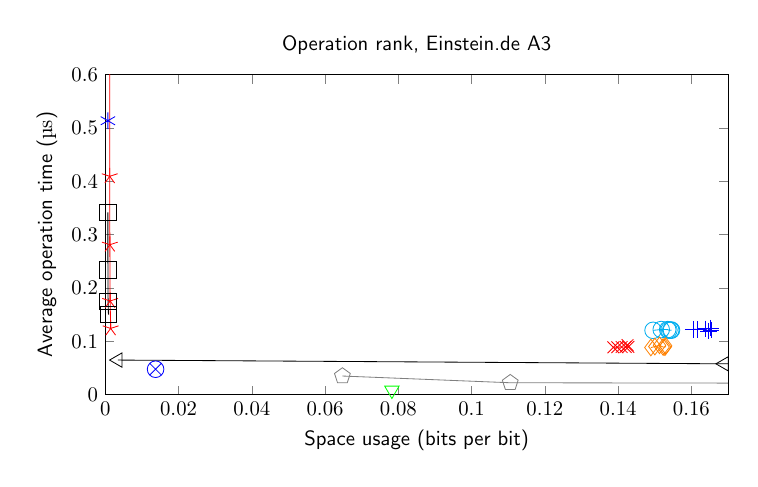}
    \includegraphics[width=0.4\textwidth]{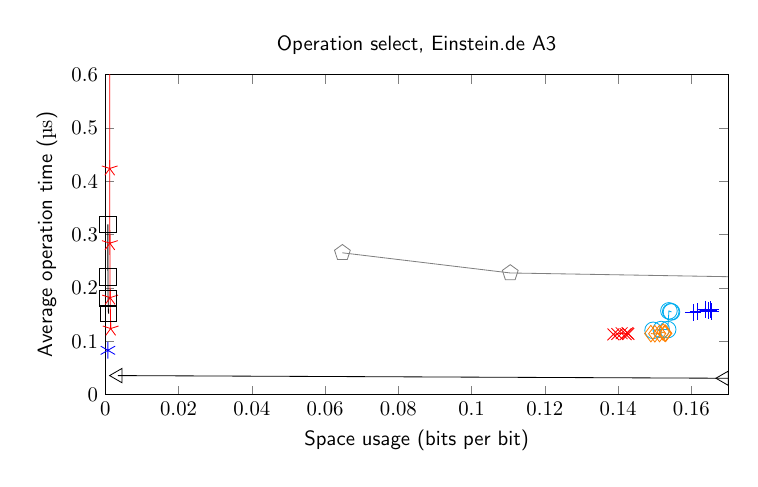}
    
    \includegraphics[width=0.4\textwidth]{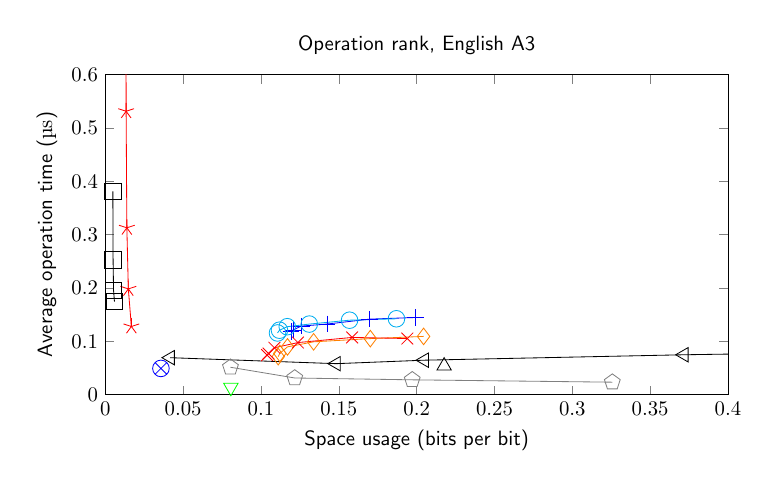}
    \includegraphics[width=0.4\textwidth]{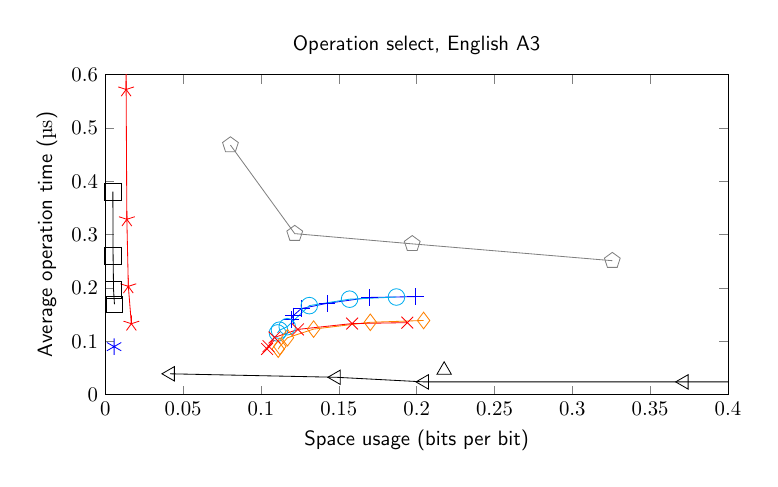}
    
    \includegraphics[width=0.4\textwidth]{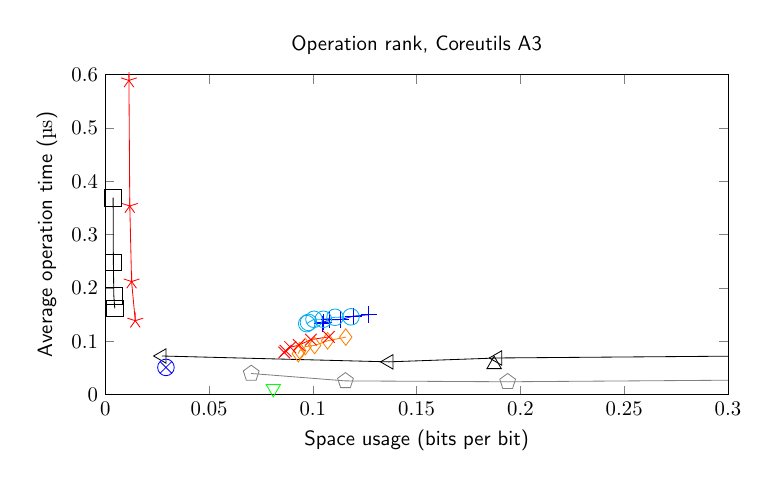}
    \includegraphics[width=0.4\textwidth]{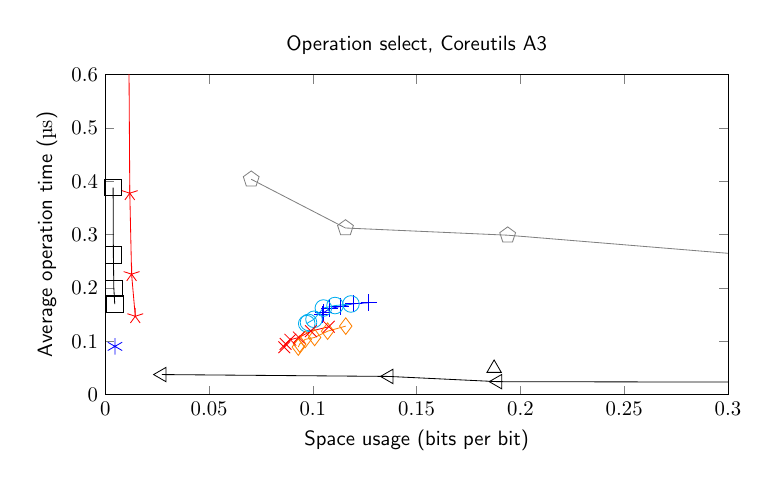}
    \includegraphics[width=0.7\textwidth]{Plots/BV2.0/leyenda_rank.pdf}
    \caption{Experimental space/time trade-offs for different compressed bit vector representations on bit sequences with runs. The bit vectors used in the tests correspond to sequences $B_0,\ldots, B_{p-1}$ using partition approach A3. Results for operation $\rank$ are shown in the left column, whereas $\select$ is shown in the right column.}
    \label{fig:rank-select-B-runs-A3}
\end{figure}

\begin{figure}[h]
    \centering
    \includegraphics[width=0.4\textwidth]{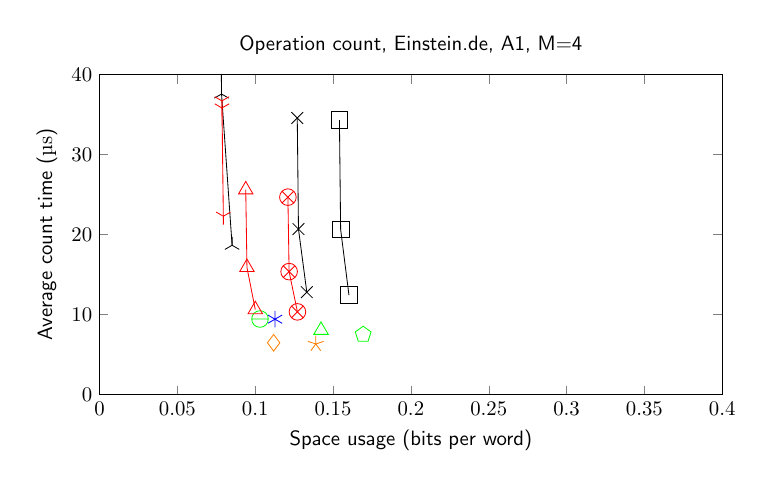}
    \includegraphics[width=0.4\textwidth]{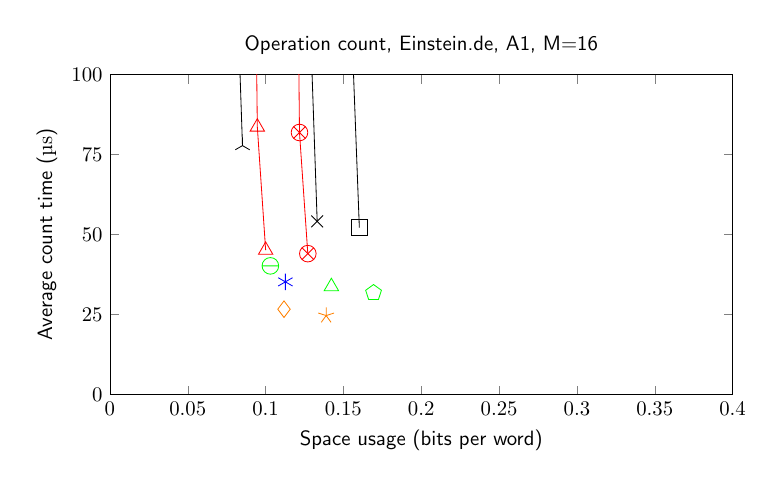}

    \includegraphics[width=0.4\textwidth]{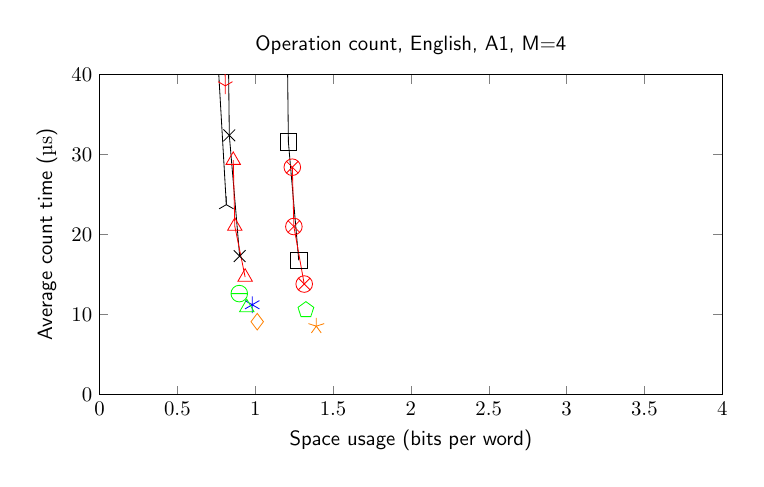}
    \includegraphics[width=0.4\textwidth]{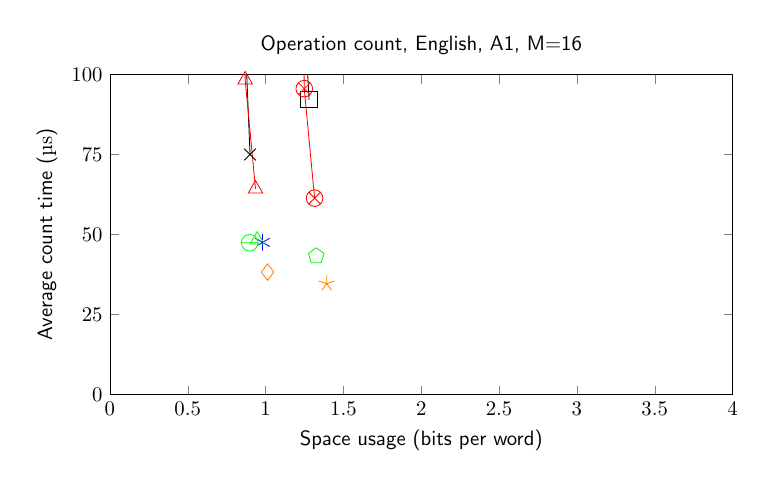}
    \includegraphics[width=0.4\textwidth]{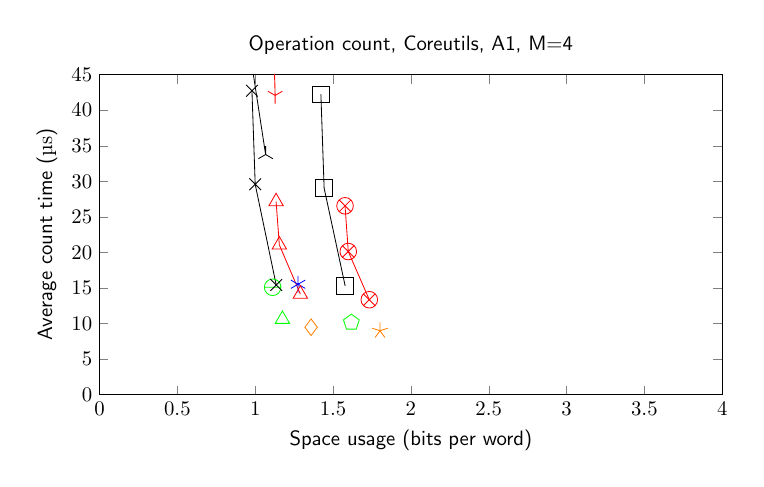}
    \includegraphics[width=0.4\textwidth]{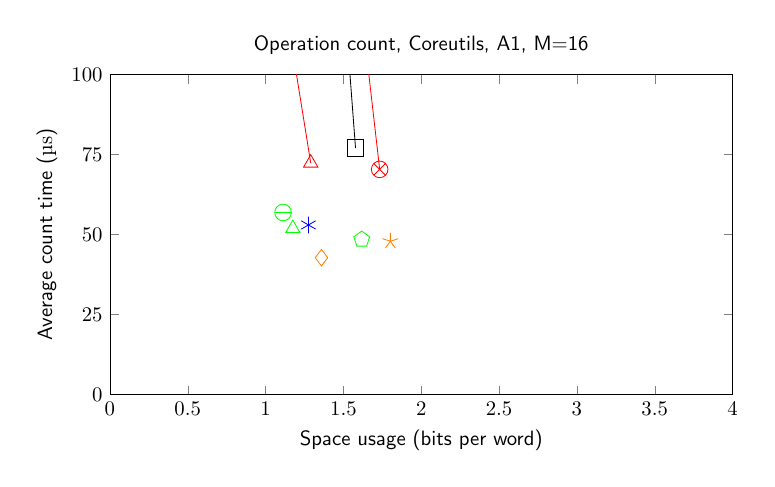}
     \includegraphics[width=0.7\textwidth]{Plots/Count-OPT/leyenda_opt.pdf}

    \caption{Space/time trade-off for count queries on the Burrows-Wheeler Transform (see Algorithm \ref{alg:backward-search}) of different large-alphabet texts, for the alphabet partition alternative A1 using patterns of length 4 and 16.}
    \label{fig:count-BWT-A1}
\end{figure}

\begin{figure}[h]
    \centering
    \includegraphics[width=0.4\textwidth]{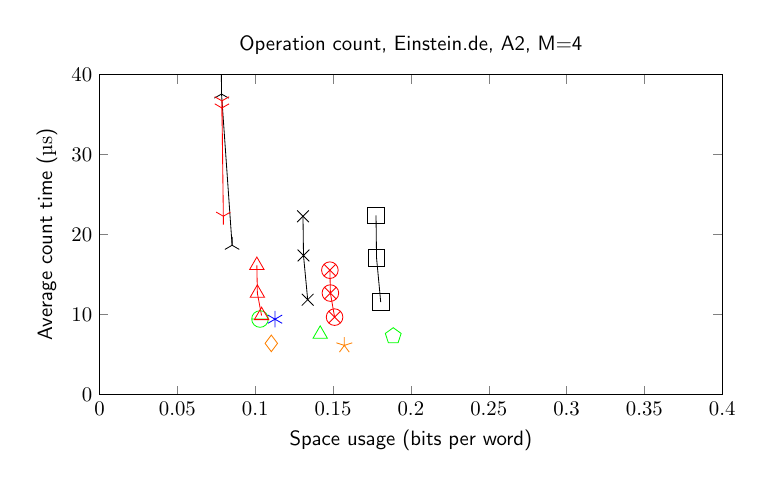}
    \includegraphics[width=0.4\textwidth]{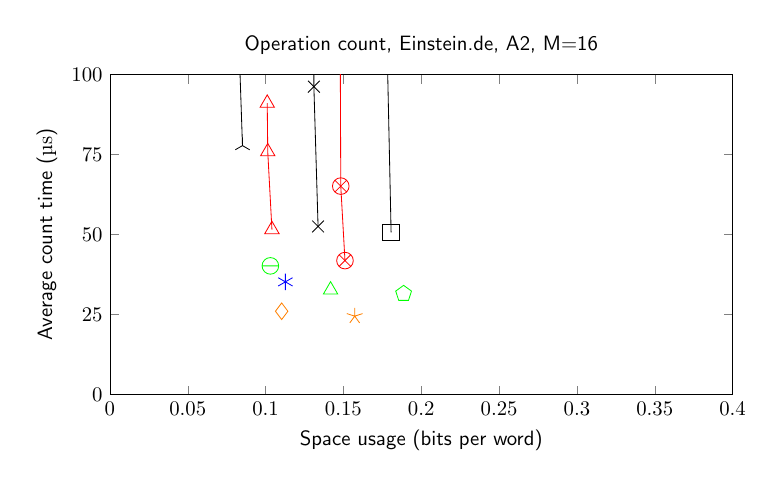}

    \includegraphics[width=0.4\textwidth]{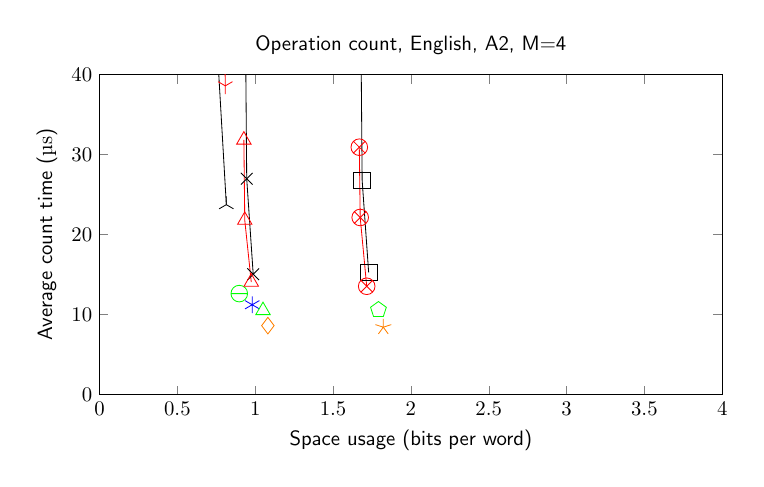}
    \includegraphics[width=0.4\textwidth]{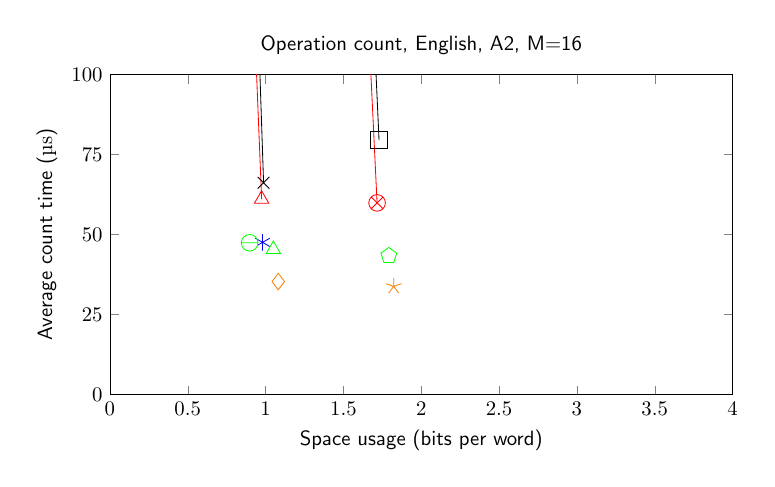}
    \includegraphics[width=0.4\textwidth]{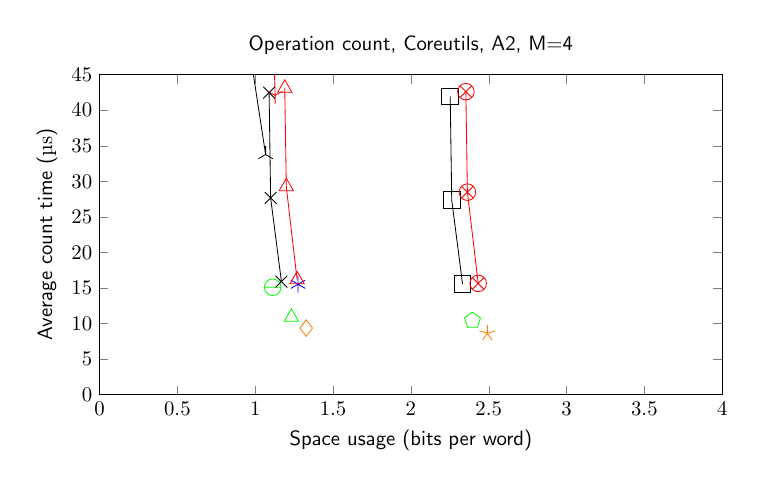}
    \includegraphics[width=0.4\textwidth]{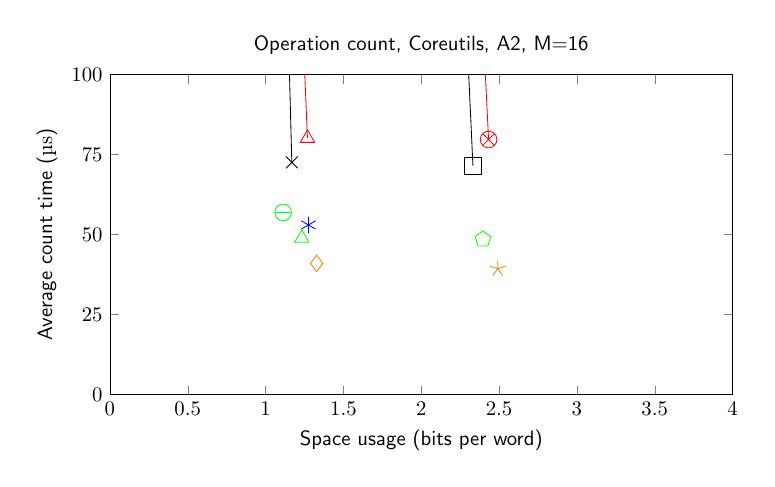}
     \includegraphics[width=0.7\textwidth]{Plots/Count-OPT/leyenda_opt.pdf}

    \caption{Space/time trade-off for count queries on the Burrows-Wheeler Transform (see Algorithm \ref{alg:backward-search}) of different large-alphabet texts, for the alphabet partition alternative A2 using patterns of length 4 and 16.}
    \label{fig:count-BWT-A2}
\end{figure}

\begin{figure}
    \centering
    \includegraphics[width=0.4\textwidth]{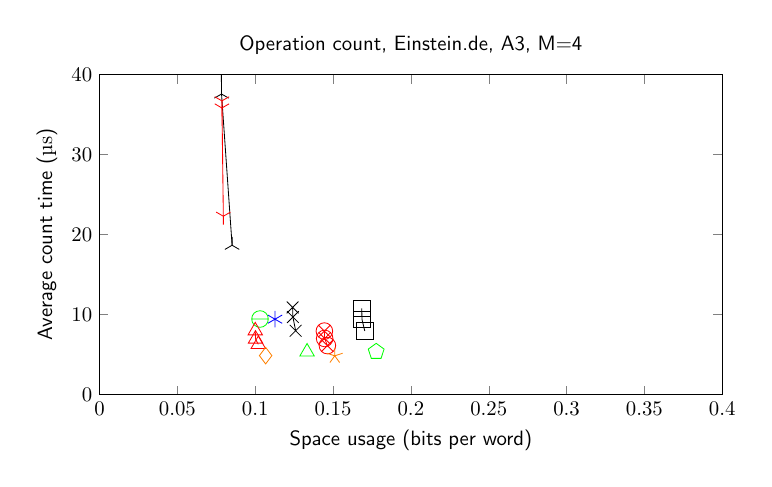}
    \includegraphics[width=0.4\textwidth]{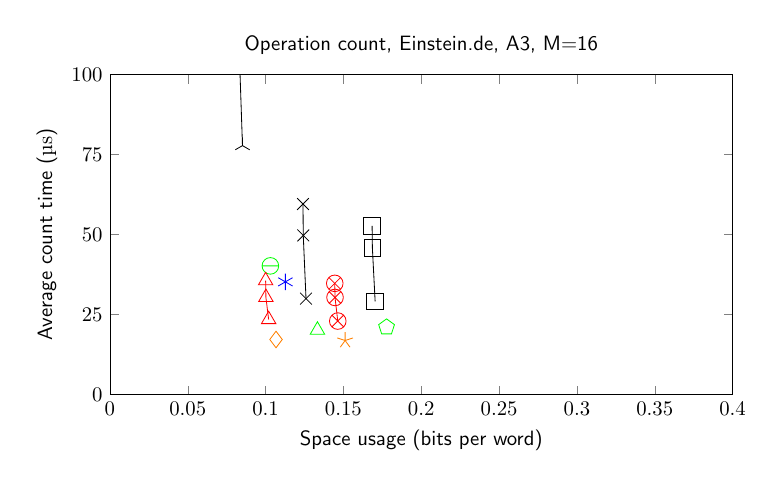}

    \includegraphics[width=0.4\textwidth]{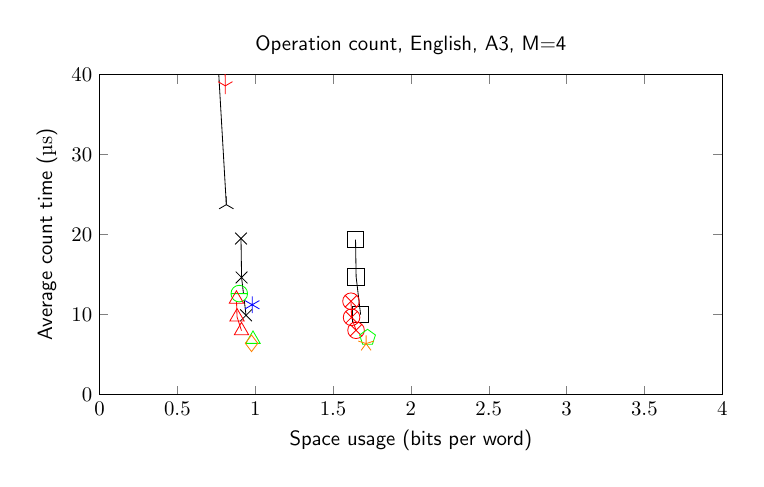}
    \includegraphics[width=0.4\textwidth]{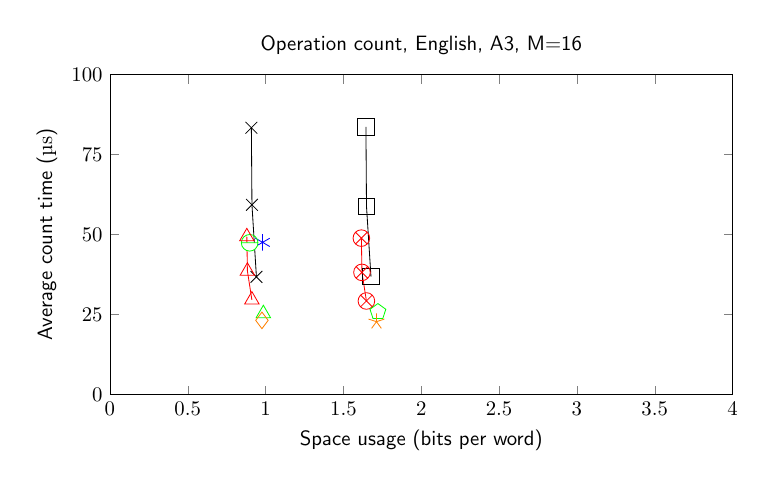}
    \includegraphics[width=0.4\textwidth]{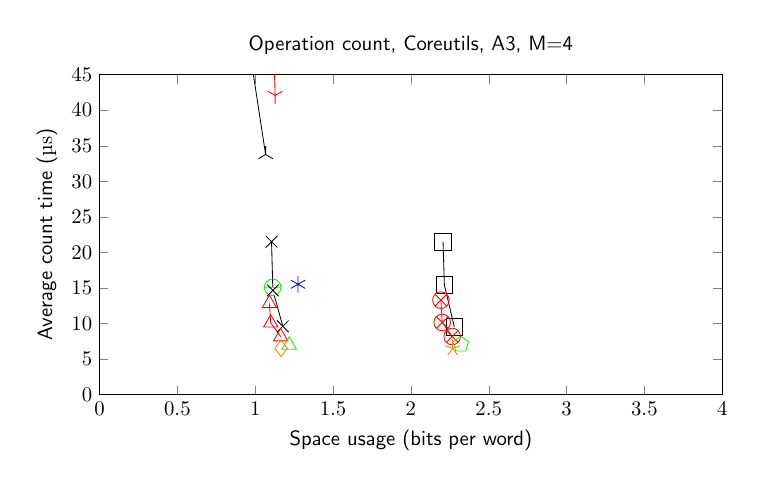}
    \includegraphics[width=0.4\textwidth]{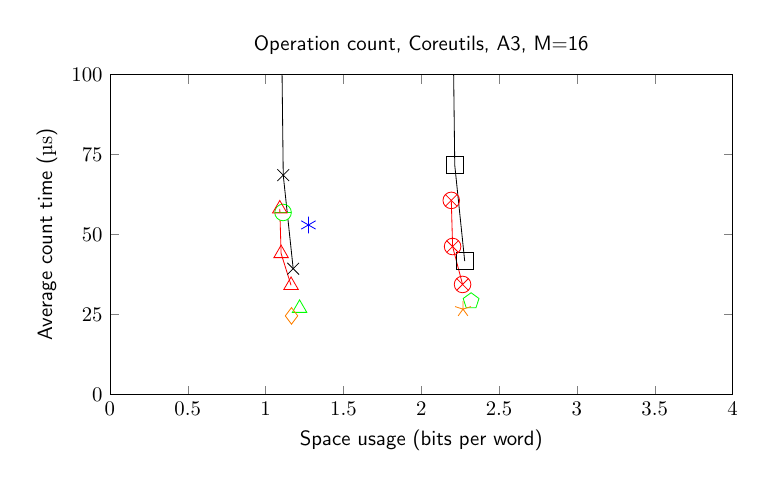}
     \includegraphics[width=0.7\textwidth]{Plots/Count-OPT/leyenda_opt.pdf}

    \caption{Space/time trade-off for count queries on the Burrows-Wheeler Transform (see Algorithm \ref{alg:backward-search}) of different large-alphabet texts, for the alphabet partition alternative A3 using patterns of length 4 and 16.}
    \label{fig:count-BWT-A3}
\end{figure}

To be comprehensive regarding our experimental results from Section \ref{sec:run-string-experiments}, in this section we show:
\begin{itemize}
    \item in Figures \ref{fig:rank-select-B-runs-A1}, \ref{fig:rank-select-B-runs-A2}, \ref{fig:rank-select-B-runs-A3}, the experimental results for operations $\rank$ and $\select$ on run-length compressed bit sequences, for partition approaches A1, A2, and A3 (compare with results for A4 in Figure \ref{fig:rank-select-B-runs}).
    
    \item in Figures \ref{fig:count-BWT-A1}, \ref{fig:count-BWT-A2}, and \ref{fig:count-BWT-A3}, the experimental results for count operation (see Section \ref{sec:experiments-RL-strings}) for alphabet partition approaches A1, A2, and A3 (compare with results for A4 in Figure \ref{fig:count-BWT-A4}). 

\end{itemize}

\end{document}